\documentclass[10pt,journal,compsoc]{IEEEtran}
\usepackage{graphicx}
\usepackage{textcomp}
\usepackage{xcolor}
\usepackage{colortbl}
\definecolor{mygray}{gray}{0.98}
\definecolor{mygray1}{gray}{0.93}
\usepackage{algorithm}  
\usepackage{algpseudocode}  
\usepackage{amsmath}  
\usepackage{amsfonts}
\usepackage{amssymb}
\usepackage{latexsym}
\usepackage{caption, subcaption}
\usepackage{enumitem}
\usepackage{booktabs}
\usepackage{listings}
\usepackage{chngpage}
\usepackage{flushend}
\usepackage{adjustbox}
\usepackage{bbding}
\usepackage{utfsym}
\usepackage{multirow}
\usepackage{textcomp,booktabs}
\usepackage{wasysym}
\definecolor{seagreen}{rgb}{0.18, 0.55, 0.34}
\definecolor{royalpurple}{rgb}{0.6, 0.4, 0.8}

\usepackage{ragged2e}
\usepackage[colorlinks,
            linkcolor=blue,
            anchorcolor=blue,
            citecolor=blue]{hyperref}

\ifCLASSOPTIONcompsoc
  \usepackage[nocompress]{cite}
\else
  \usepackage{cite}
\fi
\ifCLASSINFOpdf
\else
\fi
\begin{document}
\title{\textit{ProSecutor}: Protecting Mobile AIGC Services on Two-Layer Blockchain via Reputation and Contract Theoretic Approaches}

\author{Yinqiu Liu,
        Hongyang~Du,
        Dusit~Niyato,~\IEEEmembership{Fellow,~IEEE},~% <-this % stops a space
        Jiawen~Kang,
        Zehui~Xiong, \\
        Abbas Jamalipour,~\IEEEmembership{Fellow,~IEEE},
        and
        Xuemin (Sherman)~Shen,~\IEEEmembership{Fellow,~IEEE}
        
\IEEEcompsocitemizethanks{\IEEEcompsocthanksitem Y. Liu, H. Du, and D. Niyato are with the School of Computer Science and Engineering, Nanyang Technological University, Singapore (E-mail: yinqiu001@e.ntu.edu.sg, hongyang001@e.ntu.edu.sg, and dniyato@ntu.edu.sg)

\IEEEcompsocthanksitem J. Kang is with the School of Automation, Guangdong University of Technology, China (e-mail: kavinkang@gdut.edu.cn)

\IEEEcompsocthanksitem Z. Xiong is with the Pillar of Information Systems Technology and Design, Singapore University of Technology and Design, Singapore (e-mail: zehuixiong@sutd.edu.sg)

\IEEEcompsocthanksitem A. Jamalipour is with the School of Electrical and Information Engineering, University of Sydney, Australia (e-mail: a.jamalipour@ieee.org)

\IEEEcompsocthanksitem X. Shen is with the Department of Electrical and Computer Engineering, University of Waterloo, Canada (e-mail: sshen@uwaterloo.ca)

}% <-this % stops an unwanted space
}

\IEEEtitleabstractindextext{%
\begin{abstract}
\justifying\let\raggedright\justifying 
Mobile AI-Generated Content (AIGC) has achieved great attention in unleashing the power of generative AI and scaling the AIGC services.
By employing numerous Mobile AIGC Service Providers (MASPs), ubiquitous and low-latency AIGC services for clients can be realized. 
Nonetheless, the interactions between clients and MASPs in public mobile networks, pertaining to three key mechanisms, namely MASP selection, payment scheme, and fee-ownership transfer, are unprotected. 
In this paper, we design the above mechanisms in a systematic approach and present the first blockchain to protect mobile AIGC, called \textsf{ProSecutor}.
Specifically, by roll-up and layer-2 channels, \textsf{ProSecutor} forms a two-layer architecture, realizing tamper-proof data recording and atomic fee-ownership transfer with high resource efficiency.
Then, we present the Objective-Subjective Service Assessment ($\mathrm{OS}^2\mathrm{A}$) framework, which effectively evaluates the AIGC services by fusing the objective service quality with the reputation-based subjective experience of the service outcome (i.e., AIGC outputs). 
Deploying $\mathrm{OS}^2\mathrm{A}$ on \textsf{ProSecutor}, firstly, the MASP selection can be realized by sorting the reputation.
Afterward, the contract theory is adopted to optimize the payment scheme and help clients avoid moral hazards in mobile networks.
We implement the prototype of \textsf{ProSecutor} on BlockEmulator.
Extensive experiments demonstrate that \textsf{ProSecutor} achieves 12.5$\times$ throughput and saves 67.5\% storage resources compared with BlockEmulator.
Moreover, the effectiveness and efficiency of the proposed mechanisms are validated.
\end{abstract}

% Note that keywords are not normally used for peer review papers.
}

\maketitle

\IEEEdisplaynontitleabstractindextext

% \fi
\IEEEpeerreviewmaketitle

\IEEEraisesectionheading{\section{Introduction}\label{sec:introduction}}

\IEEEPARstart{S}{parked} by the phenomenal success of ChatGPT, AI-Generated Content (AIGC) has attracted significant attention from both industry and academia \cite{liuaigc}. 
As the latest paradigm for content creation in the Metaverse era, AIGC enables computers to generate multimedia outputs automatically (e.g., images, videos, even 3D avatars), significantly promoting generation efficiency and saving massive time and cost. 
Moreover, it also makes professional artwork creation accessible for even untrained users and stimulates people's creativity. 
From 2022 to 2023, we have witnessed the successful attempt of AIGC in various fields, such as Stable Diffusion in text-to-image generation, ChatGPT in Q \& A, and Microsoft Copilot in the daily office.
According to \textit{Acumen}, AIGC is projected to achieve a global market size of USD 110.8 Billion by 2030, growing at a compound annual growth rate of 34.3\% from 2022 to 2030 \cite{Acumen}.

\subsection{Background}
With the deepening of AIGC applications, the scalability concern is eminently exposed. 
Currently, most AIGC services rely on large pre-trained models with billions of parameters, consuming considerable storage and computation resources.
For instance, running Stable Diffusion requires at least one NVIDIA Ampere GPU with 6 GB memory \cite{liuaigc}, which is unaffordable for many resource-constrained clients \cite{verma2023survey}.
To this end, researchers recently presented the concept of \textit{Mobile AIGC} and successfully developed a series of on-device AIGC models, e.g., MediaPipe and PaLM 2-Gecko by Google \cite{Google} and the lightweight Stable Diffusion by Chen et al. \cite{chen2023speed}.
In the mobile AIGC era, clients can request AIGC inferences from Mobile AIGC Service Providers (MASPs) \cite{xu2023unleashing}.
Since MASPs are close to clients, low service latency can be realized.
Additionally, clients are able to customize the AIGC services, e.g., sharing real-time background information with MASPs to render immersive 3D environments.
Furthermore, the network-wide resources and service requests can be provisioned, forming the \textit{AIGC-as-a-Service} paradigm \cite{du2023generative}.
Despite these advantages, the interactions between clients and MASPs in mobile AIGC are complicated, pertaining to the following mechanisms.
\begin{itemize}
    \item \textbf{MASP Selection}: In mobile AIGC, each client can access multiple nearby MASPs with varying computing power, capability, and reliability. Hence, the MASP selection mechanism should incorporate these factors and select the best MASP with the highest probability of meeting the client's service requirements. 
    \item \textbf{Payment Scheme}: Afterward, the client confirms service details with the selected MASP. A payment scheme is required, which specifies the payment method (e.g., pre-paid or post-paid) and the amount of the service fee (e.g., fixed or floating value) according to the service quality promised by the MASP.
    \item \textbf{Fee-Ownership Transfer}: Once finishing the AIGC inferences, a transfer mechanism should be employed. In this way, the client and MASP can transfer the service fee and the ownership of the AIGC output to each other in a secure manner. 
\end{itemize}

\vspace{-0.15cm}
\subsection{Motivation}
Although similar mechanisms have been studied separately in many other scenarios, the unique features of mobile AIGC bring brand-new challenges. 
Firstly, in traditional service markets, such as edge offloading, effective service provider selection can be realized by firstly modeling the Quality of Experience (QoE) from the client perspective and then selecting the service providers leading to the highest QoE \cite{du2023generative, QOE}.
However, such schemes fail to support the emerging AIGC scenario due to the following reasons.
\begin{itemize}
    \item \textbf{Multimodality}: AIGC is going beyond multimedia content generation and aiming to provide an immersive fusion of multimodal services \cite{xu2023unleashing}. However, most quantitative metrics for QoE measurement are modality-specific \cite{ACR, affective1, affective2}. Hence, we need to extend various QoE models that adapt to different AIGC modalities, which are inflexible and cannot support the ever-complicated mobile AIGC applications.
    \item \textbf{Subjectivity}: AIGC outputs can be regarded as novel digital artwork whose judgment suffers from intrinsic subjectivity. Different clients may evaluate an AIGC output from different aspects. For instance, even if an AIGC image performs well in the PyTorch Image Quality tests \cite{kastryulin2022pytorch}, it may not achieve satisfying QoE if its style (e.g., realism or abstractism) does not match the client's expectations and personal preference. 
\end{itemize} 
For the payment scheme, the clients suffer from information asymmetry in mobile AIGC \cite{IA}. 
To be specific, since the resources invested by MASPs for performing AIGC inferences are unobserved, the clients are threatened by the moral hazard \cite{moralhazard}. 
In this case, if the clients pay the fixed AIGC service fee in one lump sum, dishonest MASPs might not provide high-quality service as promised to save computation resources. 
Finally, the fee-ownership transfers in mobile AIGC are vulnerable since the anonymous clients and MASPs may repudiate without being afraid of prosecution. 
For example, clients can cancel ongoing payments immediately after receiving the AIGC output and vice versa. 
Consequently, the \textit{atomicity}, i.e., whether the operations in one transfer all occur or nothing occurs, is broken.

\subsection{Our Work and Contributions}
\begin{figure*}[tbp]
\centerline{\includegraphics[width=1.6\columnwidth]{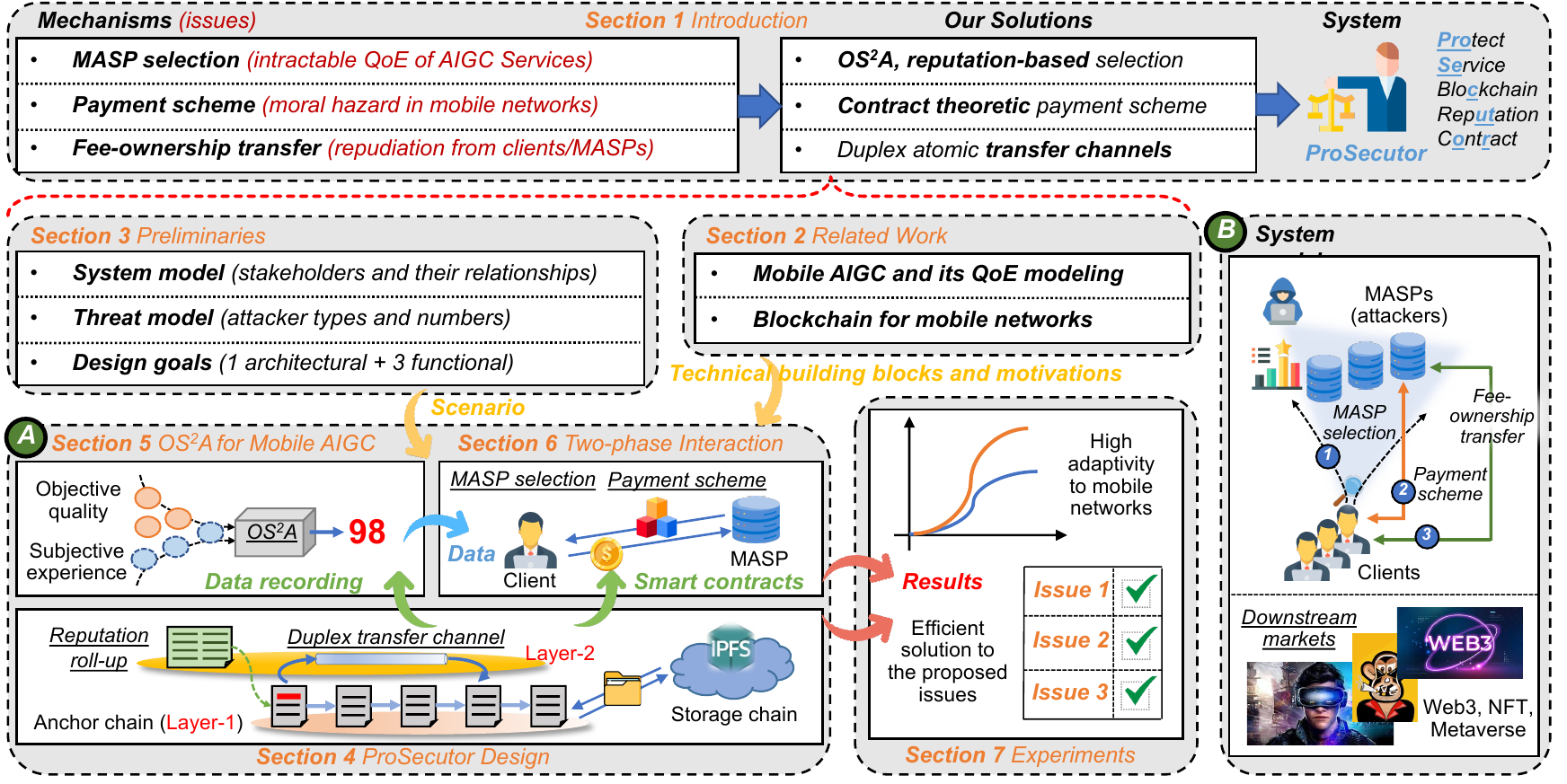}}
\caption{The structure of this paper and the system model of the mobile AIGC market.}
\vspace{-0.1cm}
\label{structure}
\end{figure*}

In this paper, we design the aforementioned mechanisms systematically to protect mobile AIGC.
Inspired by Discrete Choice Modeling (DCM) \cite{choicemodeling}, we overcome the obstacles on AIGC service evaluation by presenting a novel framework called \textit{Objective-Subjective Service Assessment ($\mathrm{OS}^2\mathrm{A}$)}.
Particularly, $\mathrm{OS}^2\mathrm{A}$ consists of both objective and subjective components, called $\mathrm{OS}^2\mathrm{A}_O$ and $\mathrm{OS}^2\mathrm{A}_S$, respectively.
As shown in Fig. \ref{structure}-Part A, the former measures the quality of the AIGC service process by objective Key Performance Indicators (KPIs), such as service latency.
On the other hand, the client's subjective opinions of the service outcome, i.e., the AIGC outputs, are captured via $\mathrm{OS}^2\mathrm{A}_S$.
To support subjectivity and multimodality, we utilize the reputation based on Multi-Weight Subjective Logic (MWSL) \cite{MWSL} to model the $\mathrm{OS}^2\mathrm{A}_S$, allowing clients to express their opinions by customized models and personal preferences. 
Leveraging $\mathrm{OS}^2\mathrm{A}$, MASP selection and payment scheme are realized.
Firstly, the clients select MASPs according to $\mathrm{OS}^2\mathrm{A}_S$, since higher reputation indicates that the MASPs are more likely to be reliable.
Afterward, the contract theory is adopted to optimize the payment scheme, which can maximize clients' utility while circumventing the moral hazard. 

Nonetheless, the proposed mechanisms are vulnerable in practical mobile AIGC.
For instance, $\mathrm{OS}^2\mathrm{A}$ is maintained based on the KPI and opinions reported by all clients, which attackers might disrupt. 
\textit{To this end, we present the first mobile AIGC-oriented blockchain, called} \textsf{ProSecutor}.
Specifically, by roll-up \cite{rollup} and layer-2 channels \cite{channel}, \textsf{ProSecutor} builds a two-layer architecture to overcome the resource limitations in mobile AIGC.
To defend against $\mathrm{OS}^2\mathrm{A}$ tampering, \textsf{ProSecutor} validates and stores all the historical data in an immutable chained ledger.
Moreover, \textsf{ProSecutor} serves as the Turing-complete smart contract engine, with which we implement MASP selection, payment scheme, and atomic fee-ownership transfers. 
The contribution of this paper can be summarized as follows.
\begin{itemize}
    \item \textbf{AIGC-Oriented Blockchain}: We propose \textsf{ProSecutor}, the first blockchain system for protecting mobile AIGC. By maintaining immutable records, \textsf{ProSecutor} enables the tamper-proof reputation calculation in public mobile networks. Moreover, it serves as the smart contract engine on which we realize the atomic fee-ownership transfers. Particularly, to facilitate mobile networks, these functions are realized in lightweight manners, i.e., reputation roll-up and transfer channels.
    \item \textbf{$\mathrm{OS}^2\mathrm{A}$ Framework}: Inspired by DCM, we present $\mathrm{OS}^2\mathrm{A}$ to evaluate the mobile AIGC services. Containing two components, $\mathrm{OS}^2\mathrm{A}$ can evaluate both the objective service process and the subjective service outcome. Moreover, we utilize the MWSL-based reputation scheme to model the subjective component, making $\mathrm{OS}^2\mathrm{A}$ compatible with multimodal AIGC.
    \item \textbf{Contract Theory for AIGC}: Deploying $\mathrm{OS}^2\mathrm{A}$ in \textsf{ProSecutor}, we realize MASP selection and payment scheme by a two-phase process. Through $\mathrm{OS}^2\mathrm{A}_S$, each client can select the best MASP with the highest reputation. Then, the contract theory is utilized to optimize the payment scheme, protecting the deserved utility of the client against moral hazard.
    \item \textbf{Experimental Results}: We develop \textsf{ProSecutor} prototype with $\mathrm{OS}^2\mathrm{A}$ framework. Three mechanisms, i.e., reputation-based MASP selection, contract theoretic payment scheme, and atomic transfer, are also developed. Extensive experiments in mobile environments validate the lightweight features of \textsf{ProSecutor} and the effectiveness of the proposed mechanisms.
\end{itemize}

\subsection{Organization}

The structure of this paper is illustrated in Fig. \ref{structure}. 
Section 2 reviews the existing related works, which not only provides technical building blocks but also states the motivations. 
Then, we model the mobile AIGC scenario in Section 3. 
Section 4 demonstrates \textsf{ProSecutor} design, offering a lightweight and secure environment for data recording and mechanism executions.
In addition, the atomic fee-ownership transfer is presented.
Section 5 illustrates the design of $\mathrm{OS}^2\mathrm{A}$ framework. 
Deploying $\mathrm{OS}^2\mathrm{A}$ on \textsf{ProSecutor}, the MASP selection and contract theoretic payment scheme are discussed in Section 6. 
Section 7 implements the prototype of \textsf{ProSecutor} and analyzes the experimental results. 
Finally, Section 8 draws the conclusion.

\section{Related Work}
In this section, we review some related works regarding mobile AIGC and its QoE modeling, as well as the progress of mobile-oriented blockchains.

\subsection{Mobile AIGC and Its QoE Modeling}
%AIGC is the third-generation paradigm for content creation, following user content generation and professional content generation \cite{xu2023unleashing}.
%Although greatly inspiring the creativity and saving human labors, 
The success of AIGC relies on large pre-trained models with billions of parameters \cite{liuaigc}.
Obviously, such a paradigm is unsuitable for numerous resource-constrained clients, hindering the further development of AIGC.
To this end, researchers have been reducing the AIGC model size and alleviating the hardware requirements.
%Some lightweight text-to-image models can smoothly run on mobile devices, such as \textit{Draw Things} app on iPhone and iPad.
%Recent research further stated that even are potential to be deployed on mobile after.
On Feb. 2023, Qualcomm AI published the world's first on-device version of Stable Diffusion, one of the most famous text-to-image AIGC models \cite{qualcomm}. 
By quantizing the model parameters from FP32 to INT8, the shrinking model can be smoothly operated on common smartphones.
Likewise, Chen et al. \cite{chen2023speed} present a series of GPU-aware optimizations for Stable Diffusion, such as flash attention and Winograd Convolution, achieving the \textit{12}-second inference latency on Samsung S23 Ultra.
SnapFusion \cite{li2023snapfusion} further reduces such latency to \textit{2} seconds by step distillation.
Nowadays, various mobile AIGC applications have been launched and widely used in practice, such as \textit{dreamer}\footnote{https://stablediffusionweb.com/app} and \textit{Draw Things}\footnote{https://drawthings.ai/}.
In the foreseeable future, AIGC will further embrace mobile networks, enjoying the easy-accessible mobile communications, computation, caching, and personalization \cite{du2023generative, xu2023unleashing}.

The QoE modeling of mobile AIGC is intractable due to the unique features of AIGC outputs as artworks.
Compared with conventional scenarios like crowdsourcing and edge caching, the QoE of mobile AIGC is greatly affected by implicit subjective user preferences.
Traditionally, such subjective QoE can be measured using under study \cite{BRADLEY199449} or analytics frameworks, e.g., absolute category rating \cite{ACR}.
However, these methods are tedious and time-consuming, without the potential for real-time QoE evaluation.
From another perspective, physiological methods aim at reflecting the perceived QoE of clients directly from their perceptual and cognitive processes, e.g., heart rate, blood pressure, and temperature \cite{QoESurvey}.
Although first-hand feelings can be acquired, clients should keep wearing specific monitors, which might be invasive and unsuitable for large-scale applications.
Recently, another emerging technique called \textit{affective computing} gained significant attention in this field.
Affective computing can measure the human perceived QoE according to the affective behaviors driven by human emotions, e.g., facial expressions, speech tone, and body gestures \cite{affective1, affective2}.
Nevertheless, training the neural networks for personalized affective behavior analysis requires tremendous time and computation resources.
Furthermore, the resulting model is modality-specific because the types of affective behaviors that clients play in different scenarios differ.
Apart from measuring implicit subjective QoE, another issue is fusing it with objective service qualities.
Du et al. \cite{du2023generative} adopt Weber–Fechner Law \cite{Weber}, which can evaluate the change of client perceivable experience caused by changing service qualities.
However, they apply the unified model for all clients while ignoring their heterogeneity in terms of personalized preference, strictness, etc.

\textbf{Insights}: In this paper, we establish a reputation scheme to measure the subjective experience of clients. 
As shown in TABLE 1, our approach outperforms existing works in two aspects.
Firstly, the clients can adopt personal QoE models and express their perceivable experiences in a uniform form of opinions. 
Note that the opinions of individual clients may contain subjectivity and bias.
Hence, WSML theory is adopted to alleviate the influence of these factors and provide each MASP with a fair reputation. 
As a result, the time-consuming user study, physical tests, and numeral network training can be circumvented.
Moreover, the existing QoE measurements are modality-specific since the user study, tests, and training objectives vary in different modalities.
In contrast, the reputation approach can be applied to any modality.
%To , we employ the DCM concept to fuse the objective service quality and the subjective experience towards the service outcome (i.e., AIGC outputs), forming the $\mathrm{OS}^2\mathrm{A}$ framework that effectively supports the MASP selection and payment scheme.
\renewcommand{\arraystretch}{1.2}
\begin{table*}[t]
\centering
\caption{The comparison of our approach with existing works. Our approach can avoid time-consuming user study and model training and overcome multimodality. Moreover, we adopt DCM to fuse the objective service quality and subjective service experience, which has not been covered previously (denoted by $\CIRCLE$).}
\begin{tabular}{m{4cm}|m{5cm}|p{1.5cm}<{\centering}p{1.5cm}<{\centering}p{2cm}<{\centering}p{1.5cm}<{\centering}}
\rowcolor[rgb]{0.90,0.90,0.90}
\toprule[1pt]
\hline
\textbf{Method} & \textbf{Description} & \textbf{Subjectivity} & \textbf{Efficiency} & \textbf{Multimodality} & \textbf{Fusion}  \\ \hline
User Study \cite{BRADLEY199449, ACR, NIMA} & Conduct large-scale user study and use analytic frameworks to process results, to train subjective aesthetics metrics & $\usym{2714}$ & $\usym{2717}$ & $\usym{2717}$ & $\CIRCLE$ \\ \cline{1-6}
Physiological Methods \cite{QoESurvey} & Measure the user experience by monitoring users' cognitive process & $\usym{2714}$ & $\usym{2717}$ & $\usym{2717}$ & $\CIRCLE$ \\ \cline{1-6}
Affective Computing \cite{affective1, affective2} & Train personalized neural networks, associating users' experiences with their emotions & $\usym{2714}$ & $\usym{2717}$ & $\usym{2717}$ & $\CIRCLE$ \\ \cline{1-6}
\textbf{Reputation (ours)} & Users adopt their own QoE models to express their opinions, which are calibrated by MWSL & $\usym{2714}$ & $\usym{2714}$ & $\usym{2714}$ & $\usym{2714}$ \\ \cline{1-6}
\bottomrule[1pt]
\end{tabular}
\label{tab:eafe}
\end{table*}
\renewcommand{\arraystretch}{1}
\subsection{Blockchain for Mobile Networks}
Two categories of research exist in the field of mobile-oriented blockchains.
Firstly, as a secure computing platform, blockchain can help mobile networks in various aspects, including key management \cite{9387145}, resource management \cite{9376903}, task offloading \cite{9933642}, mobile crowdsensing \cite{9521727}, and data/resource trading \cite{9446945}.
The general idea is to use blockchains to provide an immutable and traceable ledger and save all the sensitive historical data on-chain, thus defending the malicious tampering \cite{9387145}.
Moreover, blockchain supports Turing-complete smart contracts, which can implement any application logic and are executed automatically when pre-defined conditions are met.
Hence, the dependence on mutual trust between mobile devices can be eliminated \cite{8664132, 9634163}.

Although showing great potential in mobile deployment, traditional blockchains (e.g., Bitcoin and Ethereum) are resource-intensive \cite{8664132} since every participant performs the consensus mechanism, saves the entire ledger copy, and synchronizes the network-wide messages.
Hence, the second research direction is blockchain optimization, making it affordable for resource-constrained mobile devices.
For instance, Zaman et al. \cite{8651742} present a lightweight mobile-oriented consensus mechanism called Proof-of-Sincerity, which reduces the overall consensus costs by encouraging collaboration among devices.
Furthermore, Xu et al. \cite{Xupaper} jointly consider the computation, storage, and communication resources consumed by blockchain and propose the resource-efficient consensus mechanism, ledger pruning, and fast synchronization, respectively. 
The emerging sharding technique has also been used in mobile blockchain, alleviating the overhead on each device \cite{9119383}.
These methods aim to refine the architectural design of blockchains, called \textit{Layer-1} methods.
Accordingly, the optimizations by building additional infrastructures, such as micropayment channels \cite{9500288}, atop blockchains are \textit{Layer-2} methods.
Finally, assisted by mobile edge computing, the computation-intensive tasks of mobile devices can be offloaded to the edge servers with sufficient physical resources \cite{8933071, 9248638}.

\textbf{Insights}: Motivated by this, we pay great attention to optimizing the architecture of \textsf{ProSecutor}, making it adapt to the mobile environment when meeting the functional goals. 
Hence, we propose the reputation roll-up and the duplex transfer channels to improve the efficiency of the two most frequently executed operations in \textsf{ProSecutor}, namely reputation update and fee-ownership transfer. 

\section{Preliminaries}
In this section, we describe the preliminaries, including the system and threat models of the mobile AIGC and the corresponding design goals of \textsf{ProSecutor}.

\subsection{System Model: Mobile AIGC}
We consider a mobile AIGC market consisting of $N$ clients and $K$ MASPs, as shown in Fig. \ref{structure}-Part B.
Let $\boldsymbol{C}$ = \{$C_1$, $C_2$, $\dots$, $C_N$\} denote the set of clients with $C_n$ being the $n^{th}$ client and $\boldsymbol{M}$ = \{$M_1$, $M_2$, $\dots$, $M_K$\} denote the set of MASPs with $M_k$ being the $k^{th}$ MASP.
Within $K$ MASPs, at most $Q$ are malicious attackers.
Let $\boldsymbol{A}$ = \{$A_1$, $A_2$, $\dots$, $A_Q$\} denote the exhaust list of all the potential attackers with $A_q$ being the $q^{th}$ attacker.

Given the resource constraints, clients can hardly generate AIGC outputs locally \cite{8761998}.
Instead, with abundant physical resources and professionality to operate AIGC models, the MASPs using mobile-edge servers, can perform AIGC inferences for clients according to the given prompts. 
Accordingly, they make profits by leasing their computation power to clients.
The attackers may launch malicious attacks to compromise security.
Without disturbance from attackers, the interaction between clients and MASPs in each round of mobile AIGC service involves three mechanisms, namely MASP selection, payment scheme, and fee-ownership transfer (see Fig. \ref{structure}-Part B).
%\begin{itemize}
%    \item \textbf{MASP Selection}: Firstly, clients select the best MASP among various candidates with different attributes, e.g., capability, workload, and reliability. 
%    \item \textbf{Payment Scheme}: Then, the two parties confirm the service details, including the requirements for the AIGC output and the corresponding payment scheme.
%    \item \textbf{Fee-Ownership Transfer}: Finally, MASPs perform the inferences and send the generated AIGC output to clients. Then, clients should pay the pre-confirmed service fee to MASP as long as the received AIGC output meets the pre-confirmed requirements.
%\end{itemize}

\subsection{Threat Model}
We consider a probabilistic polynomial-time Byzantine adversary \cite{KAD} in public mobile networks to corrupt MASPs as attackers. 
Attackers can attempt to destroy \textsf{ProSecutor} system in arbitrary manners.
In the following sections, we will discuss eight potential attack strategies and the corresponding defenses of \textsf{ProSecutor}.
We then make two assumptions: 1) at most 50\% of MASPs can be attackers at any moment, which is the prerequisite for blockchain-empowered security \cite{Xupaper}, and 2) the attackers are computationally bounded and cannot defeat the modern cryptography algorithms in polynomial time \cite{8418625}. 
Note that both the MASPs and clients might be dishonest in pursuing higher profits.
For example, since the behaviors of MASPs are hidden from clients, the MASPs may intentionally reduce the resources invested in AIGC inferences after accepting the requests, called the moral hazard effect.
Likewise, clients may misbehave by canceling the payment immediately after receiving the AIGC outputs.
Last but not least, we consider a practical mobile network where the MASPs and clients are anonymous to each other, without mutual trust \cite{9298846}.

\subsection{Design Goals}
Given the above system and threat models, we summarize four design goals for \textsf{ProSecutor} to protect mobile AIGC. 
\begin{itemize}
    \item \textbf{High Adaptability}: Before the functional goals, \textsf{ProSecutor} should first fit the mobile AIGC environment, with resource limitations, high workload, and the requirement for processing large AIGC outputs (\textbf{G1}).
    \item \textbf{Efficient MASP Selection}: The quality and experience of multimodal mobile AIGC services should be measured rationally and efficiently, thereby enabling each client to select the best MASP (\textbf{G2}). 
    \item \textbf{Optimal Payment Scheme}: The clients should produce appropriate payment schemes with MASPs according to the service configurations. Moreover, the moral hazard should not damage the clients' profits (\textbf{G3}).
    \item \textbf{Atomic Fee-ownership Transfer}: The execution of the fee-ownership transfers should be atomic, i.e., either the client and MASP receive the output ownership and service fee respectively, or the entire transfer is canceled, without intermediate possibility (\textbf{G4}).
\end{itemize}

\section{ProSecutor Design}
In this section, we describe the \textsf{ProSecutor} design.
We start with the architecture overview, including parallel chains and data structure.
Then, we describe the layer-2 design of \textsf{ProSecutor} to support reputation calculation and atomic fee-ownership transfers in mobile AIGC, namely reputation roll-up and duplex transfer channels.

\subsection{Architecture Overview}
\subsubsection{Parallel Chains and Stakeholders} 
As shown in Fig. \ref{structure}-Part A, \textsf{ProSecutor} adopts a hierarchical architecture with two parallel chains.
The bottom one is called \textit{Anchor Chain}, whose functions are operating smart contracts and recording all the historical events happening in mobile AIGC, including reputation updates, fee-ownership transfers, etc.
MASPs serve as the full anchor chain nodes that run consensus mechanisms, save the entire ledger copy, and synchronize messages with others.
Without loss of generality, we equip the anchor chain with Delegated Proof-of-Stake (DPoS) \cite{DPOS2}.
In DPoS, every full node stakes tokens and competes for the super nodes, which take turns generating new blocks at pre-configured time intervals.
Such ordered block generations bring high throughput and resource efficiency, making \textsf{ProSecutor} suitable for mobile environments.
%Correspondingly, the main purpose of the anchor chain is to record all the historical events happening in the mobile AIGC market, including the ownership-fee transfers, the reputation update, etc.
Due to resource constraints, clients rely on MASPs to access the anchor chain.
To maintain the reputation records, super nodes also serve as the Reputation Coordinators (RCOs).
Finally, we build a public key infrastructure in \textsf{ProSecutor}.
Specifically, each participant owns a public-private key pair based on Security Hash Algorithm-256 (SHA256) \cite{8664132} for asymmetric encryption/decryption and digital signatures.
Note that the unique identity and address of every \textsf{ProSecutor} stakeholder are also represented by its public key.
%Finally, each stakeholder owns a unique address, which is represented by its public key.
%Accordingly, it also holds the matching private key, whose functions will be discussed below.
%Assisted by RCOs, a public-key infrastructure can be built, enabling clients to verify MASPs' identification by their public keys registered in RCOs.
%Meanwhile, RCOs maintain the reputation of all the MASPs.

To protect the security of mobile AIGC, in some cases, the newly generated AIGC outputs should be saved on the blockchain to defend against malicious tampering.
Given the large size of AIGC content, we further deploy the \textit{Storage Chain} to save the valuable storage capacity of MASPs. 
Using Inter-Planetary File System (IPFS) \cite{IPFS}, the storage chain can split large AIGC outputs into chunks and preserve them on professional storage nodes.
For simplicity, the storage chain does not maintain an independent ledger.
Instead, the file storage function is invoked by the anchor chain using IPFS-style HTTP requests.
To fetch any stored content, the clients should provide the corresponding storage recipient, which contains the SHA256 result of the entire content as the key.
Such a hierarchical architecture is designed to decouple the data recording and file storage.

%\subsubsection{Consensus Mechanism}
%For blockchains, the consensus mechanism is the core module and directly determines their performance.
%\textsf{Parma} can adopt various mainstream consensus mechanisms, such as Proof-of-Work, Proof-of-Stake, Delegated-Proof-of-Stake (DPoS), and Proof-of-Authorities.
%Without loss of generality, we employ DPoS \cite{DPOS2} for \textsf{Parma} in this paper.
%In DPoS, every full node locks tokens and competes to become the super nodes, which then take turns to generate new blocks at pre-configured time intervals.
%Such ordered block generations bring high throughput and energy efficiency, making DPoS suitable for mobile deployment.
\begin{figure}[tbp]
\centerline{\includegraphics[width=8cm, height=5.5cm]{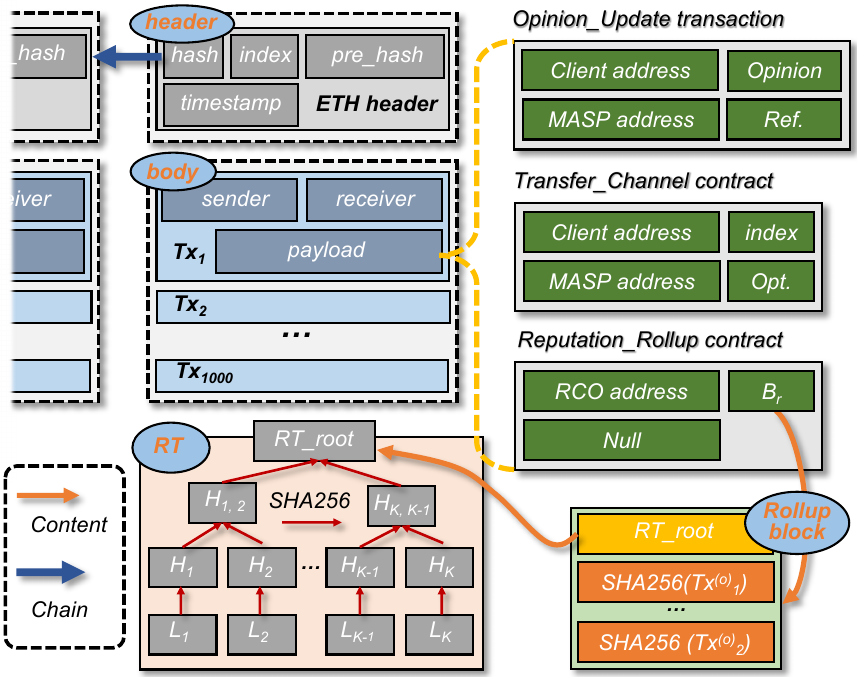}}
\caption{The data structure of \textsf{ProSecutor}. $L$ denotes the RT leaves containing MASP's address and its reputation.}
\vspace{-0.2cm}
\label{data}
\end{figure}

\subsubsection{Smart Contracts for \textsf{ProSecutor}-Aided Mobile AIGC}
With \textsf{ProSecutor}, the traditional mobile AIGC described in Section 3.1 can be extended into the blockchain-aided mobile AIGC with two features: 1) all the historical events are recorded on the anchor chain, and 2) all the operations are conducted automatically by smart contracts.
To do so, we refine the Ethereum data structure by developing new transactions and smart contracts (see Fig. \ref{data}).
\begin{itemize}
    \item \textbf{\texttt{Opinion\_\!Update} transaction}: Recall that \textsf{ProSecutor} maintains a reputation scheme for modeling $\mathrm{OS}^2\mathrm{A}_S$. Hence, we design \texttt{Opinion\_\!Update} transaction to collect the opinion of the client toward the selected MASP after each round of AIGC service. To this end, the MASP's address, the user's opinion, and the reference to specify the AIGC service should be packed. %The opinions from all clients will be calibrated by MWSL theory and form the reputation of each MASP.
    \item \textbf{\texttt{Reputation\_\!Roll-up} contract}: To support reputation roll-up, we design \texttt{Reputation\_\!Roll-up} contracts, which submit the roll-up block with thousands of opinion update records to the anchor chain and update the reputation of each MASP.
    \item \textbf{\texttt{Transfer\_\!Channel} contract}: The transfer channels are maintained by \texttt{Transfer\_\!Channel} contracts. Accordingly, \textit{index} indicates the channel index; \textit{Opt.} refers to the channel operations on the anchor chain, including \texttt{establish} and \texttt{close}. 
\end{itemize}
In the following sections, we illustrate the details of how the data structure supports the proposed mechanisms.

%\subsubsection{Data Structure}
%Fig. \ref{data} shows the data structure of \textsf{Parma} blocks and transactions.
%Specifically, each block consists of two parts, namely header and body.
%The block header saves some metadata, including \texttt{block\_height} (the index of the block), \texttt{timestamp} (block generation time), \texttt{hash} (the hash result of the block content), \texttt{pre\_hash} (the \texttt{hash} of the previous block), and \texttt{MHT\_root} (the root of the Merkle Hash Tree \cite{MHT} constructed by all the transactions in this block).

%Block body carries hundreds of recent \textsf{Parma} transactions.
%We define three types of transactions for different purposes, namely \texttt{Opinion\_\!Update}, \texttt{Reputation\_\!Roll-up}, and \texttt{Transfer\_\!Channel}.
%The basic structure of transactions is composed of three key-value pairs, i.e., [\texttt{sender}: \textit{PubKey}, \texttt{receiver}: \textit{PubKey}, \texttt{payload}: ($*$)], whose first two elements are general and specify the identities of the involved parties.
%\texttt{payload}, however, is customized according to the specific type, which will be illustrated below.

\subsection{Reputation Roll-up}
%As a distributed computing paradigm, blockchain suffers from insufficient performance and low resource efficiency.
%Taking the two most famous blockchains, i.e., Bitcoin and Ethereum as examples, their throughput is 7 and 20 Transactions Per Second (TPS), respectively, while causing huge energy waste.
%To this end, \textsf{Parma} adopts the emerging Layer-2 architecture to scale out capability and facilitate the mobile deployment.
%As mentioned in Section 1, the calculation of QoAS, especially reputation, relies on the secure and timely collection of all MASPs' historical operation records.
%This is because we can trace the reputation trend and defend the malicious tampering only if all the historical data is immutable and accessible.
%However, considering that the reputation is updated very frequently, massive records will be generated and saved on chain, which to deal with AIGC tradings.
%Therefore, we conduct the reputation roll-up to pack and compress historical reputation records.
%Recall that $\mathrm{OS}^2\mathrm{E}_S$ is reflected by the MASPs' reputation, which belongs to time series data.
%To trace reputation trends and defend tampering, all the historical reputation records should be preserved in an immutable manner.
Next, we show the layer-2 design of the anchor chain, including reputation roll-up and duplex transfer channels.
Traditionally, all the historical opinions should be saved on the ledger of each MASP. 
Nevertheless, the explosively increasing data volume wastes considerable storage resources of MASPs.
Given that opinions only serve as evidence for reputation tracing, we intend to offload them from the anchor chain and only keep the most critical bookkeeping messages.
As shown in \textbf{Algorithm 1}, we develop layer-2 reputation roll-up, containing the following steps.

\textit{1) Reputation Collection}:
%Recall that in $\mathrm{OS}^2\mathrm{A}$, the reputation of each MASP is calculated using the opinions of all clients over the AIGC market (the calculation process is discussed in Section 5.3).
Suppose that each client $C_i$ can access a nearby RCO, denoted by $RCO_i$, to update its opinion $o$ towards the selected MASP $M_i$.
$o$ is signed by $C_i$'s private key for protecting data integration and carried by the \texttt{payload} of an \texttt{Opinion\_\!Update} transaction $Tx^{(o)}$, whose \texttt{sender} and \texttt{receiver} are $C_i$ and $M_i$, respectively.
Traditionally, $RCO_i$ sends $Tx^{(o)}$ to the anchor chain, where multiple full nodes check the signature of $o$ using $C_i$'s public key.
If passed, $Tx^{(o)}$ can be packed into one future block and saved on the ledger.
With roll-up, however, the broadcast of $Tx^{(o)}$ is limited within RCOs, which maintain small-scale peer-to-peer communications to synchronize information.
For each RCO, it validates and executes $Tx^{(o)}$ locally and acquires the results, i.e., the updated reputation of $M_i$. 
RCO repeats such operations to all the coming \texttt{Opinion\_\!Update} transactions until reaching the pre-defined time or number threshold. 

\textit{2) Reputation Compression}:
When reaching the threshold, RCOs take turns compressing the received transactions.
Specifically, these transactions undergo the SHA256 operation sequentially in chronological order.
Then, a roll-up block $\mathcal{B}_r$ can be created by only containing the hashes, as shown in Fig. \ref{data}.
Compared with one block containing 1000 transactions, which typically occupies 500 Bytes \cite{8664132}, one $\mathcal{B}_r$ containing 1000 hashes only occupies 32.5 Bytes because each SHA256 output takes 256 bits \cite{9858023}.
Consequently, the data volume consumed for saving historical reputation records can be effectively compressed.

\textit{3) State Update}:
In \textsf{ProSecutor}, the reputation state of all the MASPs is organized by a binary tree called the Reputation Tree (RT).
As shown in Fig. \ref{data}, each RT leaf is formed by a MASP-reputation pair.
Then, SHA256 operations can be conducted recursively from leaves to the top until acquiring the RT root.
With the updated RT, the RCO in duty inserts the RT root into $\mathcal{B}_r$ and sends a \texttt{Reputation\_\!Roll-up} transaction $Tx^{(r)}$ to the anchor chain.
The \texttt{sender}, \texttt{receiver}, and \texttt{payload} of $Tx^{(r)}$ are its address, null, and $\mathcal{B}_r$, respectively.
All RCOs validate $\mathcal{B}_r$ by trying to reconstruct the RT based on the transactions relayed by $RCO_i$.
If passed, $Tx^{(r)}$ can be saved on-chain, with $\mathcal{B}_r$ as the bookkeeping massage.
Otherwise, super nodes can launch a poison transaction \cite{potrans} to claim the invalidity of $Tx^{(r)}$ and roll back RT to the last state.
%Since hash operation is a one-way function, attackers cannot tamper $B_r$ by forging or modifying the reputation records.
%As for reputation tracing, the clients can simply fetch the hashes that is required for reputation tracing while greatly saving the storage costs.
\begin{algorithm}[tpb]
\footnotesize \caption{Reputation Roll-up (from $RCO_i$ perspective)}
\begin{algorithmic}[1]
\Require  
$\mathbb{O}$ := \{$o_1, \dots, o_n$\} \textit{\#\#\,opinions from clients}
\Ensure 
$\mathcal{RT}^*$ \textit{\#\#\,updated RT state}
\Procedure{Reputation Collection}{$\mathbb{O}$} 
%\If{\textsf{Child\,(}$\mathbb{G}, V$\textsf{)} = $\varnothing$}
%\State \Return $V$
%\Else
\State Initialize $\mathbb{T}$ = \{\} \textit{\#\#} \texttt{Opinion\_Update} \textit{transactions}
\ForAll{$o_i \;\in\,\mathbb{O}$}
%\ForAll{$v_{ij}\;\in$ \textsf{Subtree\,(}$\mathbb{G}, v_i$\textsf{)}}
\State $Tx^{(o)}_i$.\texttt{sender} = client.\textit{PubKey}
\State $Tx^{(o)}_i$.\texttt{receiver} = MASP.\textit{PubKey}
\State $Tx^{(o)}_i$.\texttt{payload} = $o_i$
\State $\mathbb{T}$.append($Tx^{(o)}_i$)
\EndFor
\State Send $\mathbb{T}$ to other RCOs
%\EndFor
%\EndIf
\EndProcedure
\Statex
\Procedure{Reputation Compression}{$\mathbb{T}$}
\State Initiate $\mathcal{B}_r$ = \{\} \textit{\#\#} \textit{The roll-up block}
\ForAll{$Tx_i^o \in \mathbb{T}$}
\State $h_i = SHA256(Tx_i^{(o)})$
\State $\mathcal{B}_r$.append($h_i$)
\EndFor
\EndProcedure
\Statex
\Procedure{State Update}{$\mathcal{B}_r$}
\State Update RT leaves using $\mathcal{B}_r$
\State Recreate RT according to Section 4.2-3), get $\mathcal{RT}^{*}$
\State $\mathcal{B}_r$.append($\mathcal{RT}^{*}$.root) 
\State Send \texttt{Reputation\_Roll-up} transaction $Tx^{(r)}$
\State $Tx^{(r)}$.\texttt{sender} = self.\textit{PubKey}
\State $Tx^{(r)}$.\texttt{receiver} = \textit{null}
\State $Tx^{(r)}$.\texttt{payload} = $\mathcal{B}_r$
\State Send $Tx^r$ to the anchor chain for validation and saving
\EndProcedure
\end{algorithmic}
\end{algorithm}
\begin{figure}[tbp]
\centerline{\includegraphics[width=0.925\columnwidth]{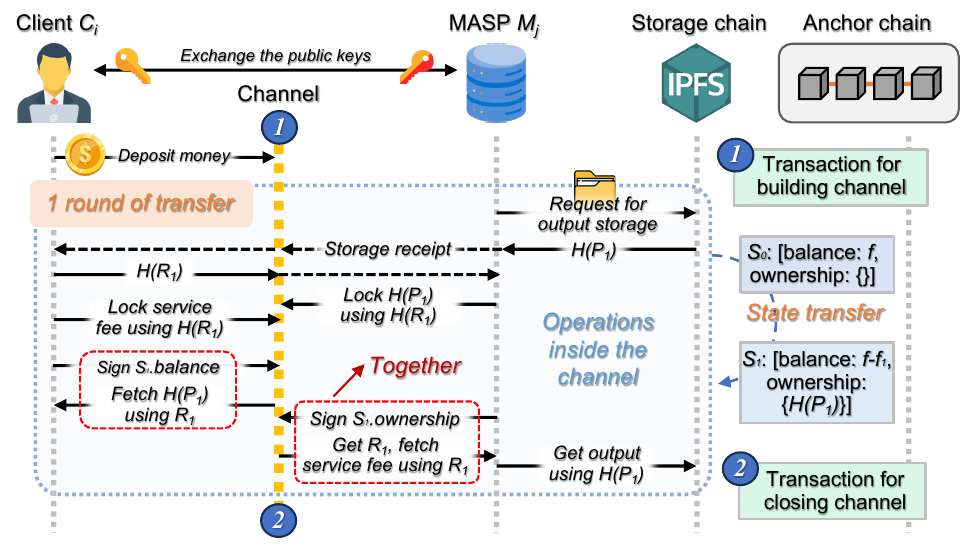}}
\caption{Illustration of atomic fee-ownership transfers. The operations framed by red dotted lines construct one atomic operation that should be executed simultaneously.}
\label{chan}
\vspace{-0.1cm}
\end{figure}

\subsection{Duplex Transfer Channel}
The second layer-2 design is duplex transfer channels between each MASP-client pair, with which we realize the atomic fee-ownership transfers.
These channels are virtual and instantiated by the specific smart contract.
Within the channel, the participants can conduct multiple rounds of atomic transfers protected by the Hash Lock (HL) protocol.
Only the channel initialization and closing need to be recorded on the anchor chain.
Since the transfers happen inside channels, low latency can be guaranteed, and the workload of the anchor chain can also be alleviated.
Next, we introduce the procedure of atomic fee-ownership transfer on the channel.

\textit{1) Channel Initialization}:
As shown in Fig. \ref{chan}, we consider two channel participants: a client $C_i$ and a MASP $M_j$.
In the first step, they exchange the public keys and send a \texttt{Transfer\_\!Channel} transaction with \textit{Opt.} \texttt{establish} to invoke the smart contract for building a transfer channel $\mathcal{E}$.
Then, $C_i$ should deposit a certain number of tokens, denoted by $f$, on $\mathcal{E}$.
Hence, the genesis state of $\mathcal{E}$ can be represented as: $\mathcal{S}_0$ := [\texttt{balance}: $f$, \texttt{ownership}: \{\}].

\textit{2) AIGC Storage}:
After the inference, $M_j$ calls the storage chain for saving the generated AIGC output $\mathcal{P}_1$.
%SC splits the content and distributes the chunks to multiple cloud servers for storage.
The receipt for storage service, as well as the key to fetch the stored content, denoted by $H(\mathcal{P}_1)$, are sent back to $M_j$.
%Based on IPFS standard, the key is the SHA256 output of the entire AIGC content, denoted by $H(C)$.
%Meanwhile, $M_j$ encrypts $H(\mathcal{P})$ using $C_i$'s public key, preventing others from stealing it.
%The receipt and encrypted $H(\mathcal{P})$ are carried by the payload of an \texttt{AIGC\_\!Storage} transaction $Tx^a$.
%Finally, $Tx^a$ is sent to the anchor chain and relayed to $C_i$ and $M_j$.

\textit{3) Fee-Ownership Transfer}:
We equip $\mathcal{E}$ with an HL protocol.
Specifically, $C_i$ first generates a random number $\mathcal{R}_1$ and sends its hash $H(\mathcal{R}_1)$ to $M_j$.
$M_j$ locks $H(\mathcal{P}_1)$ on $\mathcal{E}$ using $H(\mathcal{R}_1)$, and only the MASP providing $\mathcal{R}_1$ can unlock.
Likewise, $C_i$ also locks the pre-defined service fee, denoted as $f_1$, on $\mathcal{E}$.
With $f_1$ and $H(\mathcal{P}_1)$, $\mathcal{E}$ can update the channel state to $\mathcal{S}_1$ := [\texttt{balance}: ($f - f_1$), \texttt{ownership}: \{$H(\mathcal{P}_1)$\}].
After that, the two parties can attempt to unlock the properties locked on $\mathcal{E}$.
Firstly, $C_i$ unlocks $H(\mathcal{P}_1)$ using $\mathcal{R}_1$, which exposes $\mathcal{R}_1$ to $M_j$.
Meanwhile, it should sign the updated balance, i.e., $f-f_1$, using its private key.
With $\mathcal{R}_1$, $M_j$ can then unlock the fee $f_1$, and correspondingly, it should sign the updated ownership, i.e., $H(\mathcal{P}_1)$ on $\mathcal{E}$.
Given that all the steps are written in smart contract logic, which cannot be changed after invoking, once $C_i$ successfully unlocks $H(\mathcal{P}_1)$, $M_j$ can be guaranteed to receive the service fee, without being affected by repudiation. 

\textit{4) Channel Closing}:
Step 3 illustrates one round of fee-ownership transfer.
The two parties can repeat step 3 multiple times until $C_j$ spends all its balance.
Suppose $\mathcal{E}$ is closed after $\omega$ rounds of transfers.
The smart contract for channel closing can be invoked, collecting all the signed channel states (i.e., \{$\mathcal{S}_0$, \dots, $\mathcal{S}_{\omega}$\}). 
If passing the signature validation on the anchor chain, the balance of $C_i$ and $M_j$ can be reduced and increased by $\sum_{i=1}^\omega f_i$, respectively.
Simultaneously, the output ownership that $C_i$ acquires, i.e., \{$H(\mathcal{P}_1)$, \dots, $H(\mathcal{P}_\omega$)\}, can be formally registered on the anchor chain.
Note that since each intermediate state should be signed by both $C_i$ and $M_j$, it cannot be forged and tampered with by the attackers. 

\section{$\mathrm{OS}^2\mathrm{A}$: Objective-Subjective Service Assessment for Mobile AIGC}
In this section, we elaborate on the design of the $\mathrm{OS}^2\mathrm{A}$ framework.
Firstly, inspired by the DCM principle \cite{choicemodeling}, we present the idea of $\mathrm{OS}^2\mathrm{A}$.
Then, we demonstrate the modeling of the objective service quality and subjective service experience, respectively.
\begin{figure}[tbp]
\centerline{\includegraphics[width=0.9\columnwidth]{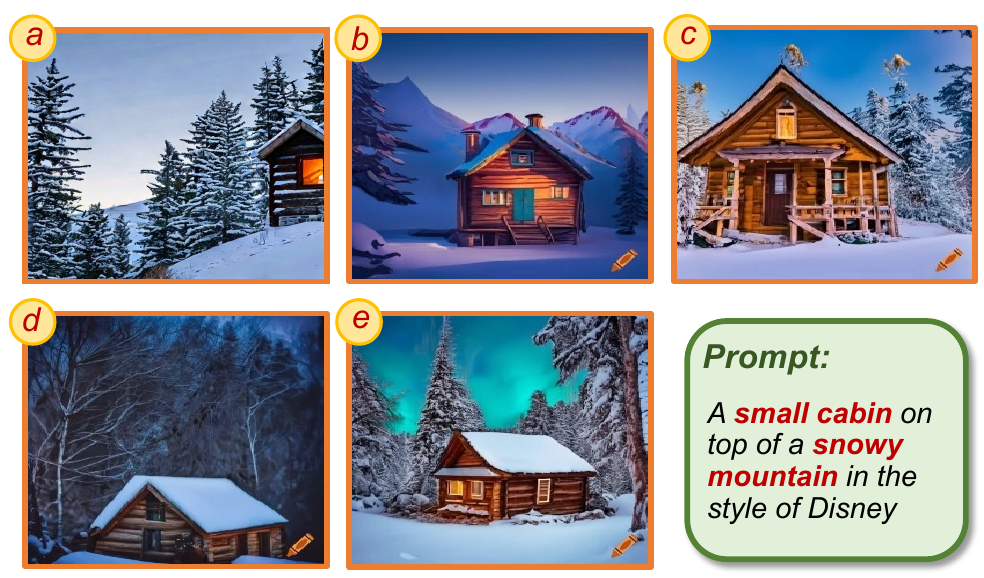}}
\caption{A series of AIGC images. Image (\textit{a}) and the others are generated by \textit{Stable Diffusion 2.1} and \textit{Craiyon V3}, respectively. The other configurations are default.}
\label{qoas}
\end{figure}

\subsection{Inspiration from DCM}
The modeling of the quality and experience of AIGC services is intractable.
For instance, Fig. \ref{qoas} shows five AIGC images generated by the same prompt.
Comparing Figs. \ref{qoas}(a) and (b), we can observe that the latter has higher ``quality" since its cabin is located in the center and contains more details.
However, from the client's perspective, the experience of Fig. \ref{qoas}(b) might be poor since it may highly prefer photorealism images like Figs. \ref{qoas}(a) and (c)-(e) rather than cartoon-styled Fig. \ref{qoas}(b).
Apart from preference, different clients may hold different standards for the AIGC services.
Some are lenient, while others might be strict. 
Back to the above example, even for Figs. \ref{qoas}(d) and (e), the strict clients might claim that the cabin is too low and there draws the unexpected aurora, respectively.
DCM, first presented by Daniel McFadden \cite{choicemodeling}, explains the cause of such situations.
This theory states that the power determining the clients' utility comes from two collaborative sources, namely objective factors and subjective factors.
The former indicates the objective attributes that clients can enjoy, while the latter is affected by the specific environment and the subjectivity of the clients themselves.
Inspired by DCM, we present the concept of \textit{Objective-Subjective Service Assessment} ($\mathrm{OS}^2\mathrm{A}$) for mobile AIGC, considering both the objective experience of the AIGC service process and the subjective experience of the AIGC outputs.
$\mathrm{OS}^2\mathrm{A}$ is defined as
\begin{equation}
   \mathrm{OS}^2\mathrm{A} = \alpha\,\Omega(\mathrm{OS}^2\mathrm{A}_O) + (1-\alpha)\,\Omega(\mathrm{OS}^2\mathrm{A}_S),
\end{equation}
where $\mathrm{OS}^2\mathrm{A}_O$ and $\mathrm{OS}^2\mathrm{A}_S$ denote the objective and subjective components, respectively.
These two items are fused linearly, with the weighting factor $\alpha$.
$\Omega(\cdot)$ is adopted to eliminate the effect of magnitude, which is defined as
\begin{equation}
    \Omega(x) = \frac{x-x_{min}}{x_{max}-x_{min}}
\end{equation}
where $x_{min}$ and $x_{max}$ are the lower and upper bounds of $x$, respectively.
In the following parts, we elaborate the calculation of $\mathrm{OS}^2\mathrm{A}_O$ and $\mathrm{OS}^2\mathrm{A}_S$.

\subsection{Objective Quality of the Service Process}
The objective quality of the AIGC service process can be evaluated by various KPIs from different perspectives. 
%Nonetheless, since many micro and latent factors can be captured by $\mathrm{OS}^2\mathrm{A}_S$, we adopt general KPIs rather than fine-grained but modality-specific ones (e.g., the image-oriented BRISQUE in \cite{du2023generative}) for measuring $\mathrm{OS}^2\mathrm{A}_O$, making it adapt to multimodal AIGC. 
Service latency is one of the most critical concerns because, given the immature AIGC governance, the currently common practice is that the client who first publishes one AIGC output owns the copyright \cite{Copy}.
Consequently, clients will always expect a service latency as small as possible to prevent similar ideas from being published by other mobile neighbors.
Without loss of generality, in this paper, we take AIGC service latency as an example to showcase the $\mathrm{OS}^2\mathrm{A}_O$ calculation.

With \textsf{ProSecutor}, the latency of each round of AIGC service comes from three sources, denoted by $\mathcal{T}_1$-$\mathcal{T}_3$.
Firstly, the selected MASP, say $M_j$, conducts AIGC inference to generate the required outputs.
After that, it calls the storage chain and uploads the generated content.
Finally, the service fee and ownership of the AIGC output should be transferred to the MASP and client, respectively.  
Denote $\mathcal{S}_p$ as the size of the AIGC output, $\Bar{b}$ as the average bandwidth, $\mathcal{D}_t$ as the difficulty of the AIGC tasks, and $\Bar{c}$ as the average amount of computational resources invested by MASP per second.
$\mathcal{T}_1$ and $\mathcal{T}_2$ are fixed and can be obtained as $\mathcal{S}_p/\Bar{b}$ and $D_t/\Bar{c}$, respectively.
In contrast, $\mathcal{T}_3$ is uncertain; the following two situations may occur: 1) If the channel between $M_j$ and the client is active, the transfer can be immediately completed, i.e., $\mathcal{T}_3$ $\approx$ 0, and 2) If the channel has not been established yet or has been closed, the MASP should first build a channel and then start the transfer process.

\subsubsection{Modeling of Channel Establishment}
To establish a channel, $M_j$ needs to invoke the corresponding smart contract by a \texttt{Transfer\_\!Channel} transaction $Tx^{(c)}$ and broadcast it over the anchor chain.
Super nodes cache $Tx^{(c)}$ in their own transaction pools, in which $Tx^c$ stays in a queue and waits to be packed by a future block.
After $Tx^{(c)}$ gets packed and saved, the channel is activated, and the two parties can conduct atomic fee-ownership transfers via the channel.
The overall latency of such processes can be modeled as \cite{Latency1}
\begin{equation}
    \mathcal{T}_3 = \frac{\mathcal{T}_b + \mathcal{T}_q}{1 - p_{fork}},
\end{equation}
where $\mathcal{T}_b$ and $\mathcal{T}_q$ refer to the broadcast and queuing latency, respectively; $p_{fork}$ indicates the probability of the fork happenings, which is 0 in DPoS-based blockchains \cite{DPOS2}.

\begin{figure}[tbp]
\centerline{\includegraphics[width=0.9\columnwidth]{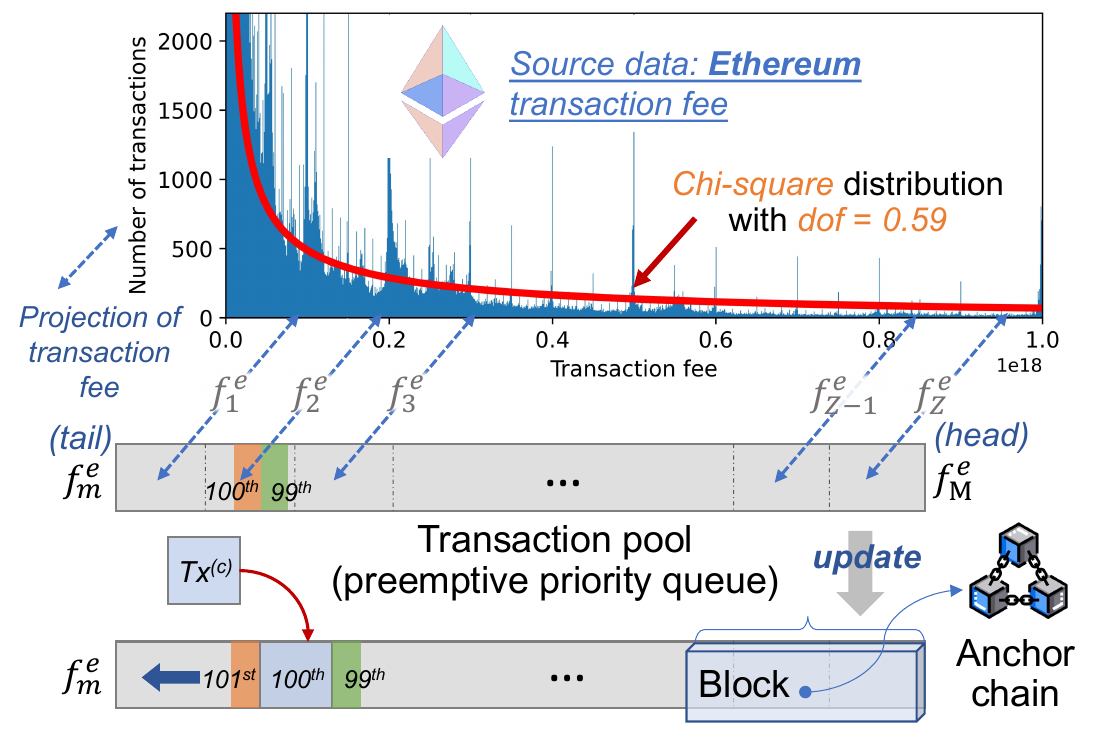}}
\caption{The illustration of transaction pool and the fitting of Ethereum transaction fee.}
\vspace{-0.4cm}
\label{pool}
\end{figure}

Suppose that \textsf{ProSecutor} uses \textit{Kademlia} protocol \cite{KAD} to implement peer-to-peer communications.
$\mathcal{T}_b$ can be defined as $\frac{log(|\mathbb{M}|)\cdot \mathcal{S}_b}{\Bar{k} \cdot \Bar{b} \cdot p_b}$, where $\mathcal{S}_b$ is the block size; $\Bar{k}$ is the average number of neighbors per MASP; $p_b$ indicates the probability of MASPs for honestly broadcasting the receiving message, which equals $1 - \frac{|\mathbb{A}|}{|\mathbb{M}|}$.
For $\mathcal{T}_q$, in \textsf{ProSecutor}, the transaction pool of each super node can be regarded as a \textit{preemptive priority queue}. 
As shown in Fig. \ref{pool}, when $Tx^{(c)}$ comes, super nodes insert it into the appropriate position according to its transaction fee, denoted by $\mathcal{F}_e$, ensuring that $\mathcal{F}_e$ of the entire queue is from high to low.
The first $\mathcal{S}_b$ transactions can be packed into the next block.
Note that $\mathcal{F}_e$ refers to the rewards that clients pay for super nodes for packing their transactions.
Such a setting aims at preventing spamming attacks \cite{8664132}, in which the attackers submit massive spamming transactions at no cost to block the system.
Suppose that the transaction pool states among super nodes are synchronized.
To model $\mathcal{T}_q$ of the given $Tx^{(c)}$, we need first to figure out what level its fee is at.
To do so, we collect one million historical real-world transactions of Ethereum\footnote{Data available at http://xblock.pro/\#/dataset/14}, whose economic system \cite{Etheconomy} is the same as \textsf{ProSecutor}.
Then, we use the Python \textit{SciPy} library to conduct fitting using various kinds of curves and find that $\mathcal{F}_e$$\sim \chi^2(0.59)$.

Suppose that all the possible $\mathcal{F}_e$ values fall into the range [$f^e_m$, $f^e_M$].
For simplicity, we split this range into $\mathcal{Z}$ parts and assume that all the transactions within the same sub-range have the same $T_q$.
As illustrated in Fig. \ref{pool}, we denote these sub-ranges by their mean, i.e., \{$f^e_1, \cdots, f^e_\mathcal{Z}$\}, satisfying $f^e_m \textless f^e_1 \textless \dots \textless f^e_\mathcal{Z} \textless f^e_M$.
Suppose that block generation in \textsf{ProSecutor} is a poison process, denoted by $Poi(\lambda t)$. $\mathcal{T}_q$ for the transaction with fee $f^e_z$ can be modelled as \cite{latency21}
\begin{equation}
\begin{split}
    \mathcal{T}_q(f^e_z) &= \frac{1}{\chi^2(f^e_1; 0.59)}\cdot \Big[\tau(\sum_{k=z}^{\mathcal{Z}}\chi^2(f^e_k; 0.59)\lambda) \\ -& \sum_{k=z}^{\mathcal{Z}-1}\mathcal{T}_q(f^e_k)\chi^2(f^e_k; 0.59)\Big], \;\;  \forall z \in [1, 2, \dots, \mathcal{Z}\!-\!1]
\end{split}
\end{equation}
and $\mathcal{T}_q(f^e_\mathcal{Z})$ = $\tau(\chi^2(f^e_\mathcal{Z}; 0.59)\lambda)$.
Inspired by Little's Law \cite{Little}, $\tau(x)$ is defined as $E(\mathcal{L})$/$x$, in which $E(\mathcal{L})$ indicates the average length of the transaction queue and can be acquired by monitoring the running states of \textsf{ProSecutor}.
Therefore, Eq. (4) can be worked out by induction from $\mathcal{T}_q(f^e_{\mathcal{Z}-1})$ to $\mathcal{T}_q(f^e_1)$. 
From Eq. (4), we can observe that $\mathcal{T}_q$ of transactions with lower $\mathcal{F}_e$ is determined by those high-priority ones.
Finally, $\chi^2(x; 0.59)$ is defined as
\begin{equation}
     \left\{
    \begin{aligned}
    &\frac{1}{2^{(\frac{\xi}{2})}\Gamma(\frac{\xi}{2})}x^{(\frac{\xi}{2}-1)}e^{\frac{x}{2}}, &x \textgreater 0 \\
    &0, & else
    \end{aligned}
    \right.
\end{equation}
where $\Gamma(\cdot)$ represents the Gamma function, i.e., $\Gamma(x)$ = $\int_{0}^{\infty}t^{x-1}e^{-t}dt$ \cite{Gamma}, and $\xi$ is degree of freedom, i.e., 0.59.

\subsubsection{Generalization to Other KPIs}
In this paper, the reciprocal of AIGC service latency, i.e., $(\sum_{i=1}^3\mathcal{T}_i)^{-1}$, is adopted to reflect $\mathrm{OS}^2\mathrm{A}_O$.
Note that other general KPIs can also be introduced, making $\mathrm{OS}^2\mathrm{A}_O$ more comprehensive.
Suppose that a series of objective KPIs are utilized, denoted by \{$\mathcal{KPI}_1$, $\dots$, $\mathcal{KPI}_n$\}.
In this case, $\mathrm{OS}^2\mathrm{A}_O$ can be calculated as
\begin{equation}
    \mathrm{OS}^2\mathrm{A}_O = \sum_{i=1}^n \omega_i \cdot S(\mathcal{KPI}_i - t_i)\cdot \mathcal{KPI}_i,
\end{equation}
where $\omega_i$ and $t_i$ indicate the weight and threshold of $\mathcal{KPI}_i$, respectively.
$S(\cdot)$ is the Heaviside step function \cite{HSF} which outputs \textbf{1} when the KPI value exceeds the corresponding threshold, and \textbf{0} otherwise.

\subsection{Subjective Experience of AIGC Outputs}
Besides the objective KPIs, clients also care about whether the generated AIGC outputs can meet their personal requirements, which we coin as the \textit{experiences} of the AIGC service outcome.
Considering the subjectivity in artwork assessment, clients evaluate the AIGC outputs using their own standards and preferences.
Nonetheless, regardless of which kind of model is used, the final output should be converted to a \textit{boolean judgment}, whose value is \textbf{0} or \textbf{1}.
Value 1 means that the client receives a satisfying AIGC output.
In contrast, value 0 indicates that the received output suffers from flaws, e.g., low resolution, distortion, and mismatched art style. 
Then, an intuitive way is to average the judgment items from all the clients and calculate the reputation score for every MASP.
However, recall that these clients hold different levels of strictness.
We should calibrate the judgment from numerous heterogeneous clients and generate a fair overall reputation for each MASP.
To reach this goal, we use WMSL \cite{MWSL} to calibrate the judgment, as shown in \textbf{Algorithm 2}.

\subsubsection{Local Opinion}
We consider a client $C_i$ and a MASP $M_j$.
Suppose that $C_i$ has ($p_{ij} + n_{ij}$) pieces of judgement to $M_j$, in which $p_{ij}$ pieces have value 1 and $n_{ij}$ pieces have value 0.
Then, the local opinion from $C_i$ to $M_j$ can be formally defined as a vector $\omega_{ij}^{loc}$ := \{$s_{ij}^{loc}, u_{ij}^{loc}, c_{ij}^{loc}$\}.
Elements $s_{ij}^{loc}$, $u_{ij}^{loc}$, and $c_{ij}^{loc}$ measure the \textbf{S}atisfying, \textbf{U}nsatisfying, and un\textbf{C}ertainty levels of $C_i$ towards $M_j$, respectively. 
They are defined as:
\begin{equation}
    \left\{
\begin{aligned}
s_{ij}^{loc} & =  (1 - c_{ij}^{loc})\, \frac{p_{ij}}{p_{ij} + n_{ij}}\\
u_{ij}^{loc} & =  (1 - c_{ij}^{loc})\, \frac{n_{ij}}{p_{ij} + n_{ij}} \\
c_{ij}^{loc} & =  \frac{1}{p_{ij} + n_{ij}}
\end{aligned}
\right.
\end{equation}
A Local opinion is the basic unit of MWSL-based reputation.
From Eq. (7), we can observe that $s_{ij}^{loc}$ and $u_{ij}^{loc}$ indicate the portions of $C_i$ for receiving satisfied and unsatisfied outputs from $M_j$, respectively.
Meanwhile, $c_{ij}$ serves as the weighting factor and adjusts the impact of $s_{ij}^{loc}$ and $u_{ij}^{loc}$ according to $C_i$'s uncertainties to $M_j$.
The higher the number of interactions between $C_i$ and $M_j$, the higher the weights of $C_i$'s opinions.
%Then, from $C_i$ perspective, the reputation of $M_j$ can be expressed as $RP_{ij}^{loc}$ := $s_{ij}$ - $\gamma c_{ij}$, called local reputation. $\gamma$ indicates the weight of the uncertainty.
\begin{algorithm}[tpb]
\footnotesize \caption{MWSL-based Reputation and MASP Selection}
\begin{algorithmic}[1]
%\Require  
%$\mathbb{G}, \Theta$, \textit{V}(starting point);\
%\Ensure 
%$\mathbb{G}^*$\,(updated local ledger state)
\Procedure{Reputation Calculation}{} 
\State \textit{\#\#\, After $C_i$ interacts with $M_j$:}
\State Update $s^{loc}_{ij}$, $u^{loc}_{ij}$, and $c^{loc}_{ij}$ based on Eq. (7)
\State Send $\omega_{ij}^{loc}$ := \{$s_{ij}^{loc}, u_{ij}^{loc}, c_{ij}^{loc}$\} to one RCO
\\
\State \textit{\#\#\, For the recevier RCO:}
\State Synchronize $\omega_{ij}^{loc}$ with others
\State Update $\mathcal{F}_{ij}^a$, $\mathcal{F}_{ij}^r$, and $\mathcal{W}_{ij}$ based on Eqs. (8) and (9)
\\
\State \textit{\#\#\, In the clients intending to calculate $M_j$'s reputation:}
\State Fetch $\mathcal{F}_{kj}^a$, $\mathcal{F}_{kj}^r$, and $\mathcal{W}_{kj}$, $\forall k \in \mathbb{C}$
\State Apply sensitivity value $\theta$ and calculate reference opinion $\omega_{\_j}^{ref}$ := \{$s_{\_j}^{ref}, u_{\_j}^{ref}, c_{\_j}^{ref}$\} based on Eq. (10)
\State Fuse local opinion and reference opinion based on Eq. (11)
\State Calculate $M_j$'s reputation based on Eq. (12)
\EndProcedure
\Statex
\Procedure{MASP Selection}{}
\State Initialize $\mathbb{M_\mathrm{AVA}}$, containing all the available MASPs 
\State Calculate their reputation as above and $\mathbb{M_\mathrm{AVA}}$ following the descending order of reputation
\ForAll{$M_i \in \mathbb{M_\mathrm{AVA}}$} 
\State Send details of the service request     \;\;\;\textit{\#\#\, First round of handshake}
\If{$M_i$ agrees and sends back \textit{acknowledgement}}
\State Send the prompts (or the receipt and key for fetching prompts from the storage chain)
\If{$M_i$ sends back \textit{acknowledgement}}  \;\;\;\textit{\#\#\, Second round of handshake}
\State The MASP selection is successful
\Else 
\State Try the next MASP in $\mathbb{M_\mathrm{AVA}}$
\EndIf
\Else
\State Try the next MASP in $\mathbb{M_\mathrm{AVA}}$
\EndIf
\EndFor
\EndProcedure
\end{algorithmic}
\end{algorithm}

\subsubsection{Reference Opinion and Calibration}
For $C_i$, the local opinion $\omega_{ij}$ represents its personal rating to $M_j$'s services.
Directly adopting $\omega_{ij}^{loc}$ in MASP selection is possible, while several issues exist.
Firstly, $\omega_{ij}^{loc}$, from only one client, can be biased.
Secondly, $C_i$ needs to accumulate the knowledge of every newly coming MASP by interacting with it several times, which is obviously inefficient. 
Finally, the attackers might launch Sybil attacks \cite{8664132} to $C_i$, i.e., maintaining a high reputation initially to gain trust and misbehaving later.
To this end, we establish a network-wide opinion-sharing scheme, where every client shares its latest opinion towards all the MASPs that it has interacted with.
Specifically, the shared opinions are packed by RCOs using \texttt{Opinion\_\!Update} transactions, whose hashes are registered in the \texttt{Reputation\_\!Roll-up} transactions and preserved permanently on the anchor chain for tracing.

From the perspective of $C_i$, the opinions towards $M_j$ from other clients $C_k$ ($k \neq i$), are called reference opinions.
Given that these clients are heterogeneous regarding strictness, familiarity with MASPs, and knowledge of the AIGC market, their opinions should be calibrated.
We consider the client $C_k$ and MASP $M_j$.
After RCO receives $\omega_{kj}^{loc}$, it immediately calculates the following factors.

\textit{1) Familiarity}:
Intuitively, the more times one client has interacted with $M_j$, the more prior knowledge it can accumulate. 
Correspondingly, a higher weight should be assigned to the reference opinions shared by the clients who are familiar with $M_j$.
The familiarity between $C_k$ and $M_j$ is defined as
\begin{equation}
    \mathcal{F}_{kj}^a = \frac{p_{kj} + n_{kj}}{\sum_{c \in \mathbb{C}}\, (p_{cj} + n_{cj})}, 
\end{equation}
whose denominator represents the total number of times that $M_j$ conducts AIGC inferences for clients.

\textit{2) Freshness}:
In mobile AIGC, the performance of MASPs dramatically changes over time due to the rapid evolution of AIGC models, the real-time workload, the malicious attacks, etc.
Therefore, recent opinions should have a higher weight than the stale ones.
The freshness of $\omega_{kj}^{loc}$ is defined as $\mathcal{F}_{kj}^r$ = $\theta^\Delta$, $\theta \in (0, 1)$.
$\Delta$ indicates the ``age" of $\omega_{kj}^{loc}$ and is measured by $(\mathcal{I}_\alpha - \mathcal{I}_\beta)\, \frac{1}{\lambda}$, where $\mathcal{I}_\alpha$ and $\mathcal{I}_\beta$ mean the indexes of the block carrying $\omega_{kj}^{loc}$ and the latest block, respectively; $\frac{1}{\lambda}$ denotes the block generation interval in DPoS. 

\textit{3) Market Worth}:
If only considering familiarity and freshness when weighing opinions, attackers can easily launch dusting attacks \cite{8666834} to destroy the reputation.
Specifically, they can submit massive fake opinions in a short time to maliciously decrease/increase the reputation of some MASPs.
To this end, we employ the third weighting factor, called market worth and defined as
\begin{equation}
    \mathcal{W}_{kj} = \sum_{q=1}^{p_{kj} + n_{kj} - 1} \mathcal{V}(\omega_{kj}^{loc}[q]) \cdot \mathcal{F}_{kj}^r[q].
\end{equation}
Recall that $C_i$ sends the updated $\omega_{kj}^{loc}$ to RCO once it completes one round of interaction with $M_j$.
The historical opinions are preserved by RCO and are denoted as \{$\omega_{kj}^{loc}[1], \dots, \omega_{kj}^{loc}[(p_{kj} + n_{kj} - 1)]$\}.
In Eq. (9), $\mathcal{V}$ denotes the value function, which fetches the market worth of $\omega_{kj}^{loc}[q]$, i.e., the amount of service fee that $C_i$ has paid for $M_j$ in the $q^{th}$ round of AIGC service.
Correspondingly, $\mathcal{F}_{kj}^r[q]$ means the freshness of $\omega_{kj}^{loc}[q]$, which can prevent attackers from placing a large-value service order and acquiring high $\mathcal{W}_{kj}$ permanently.
%Finally, all the opinion updates in \textsf{Parma} should refer to the corresponding \texttt{Exchange} transaction.
%Hence, the attackers cannot forge market worth for their opinions.
%The detailed security analysis will be given in Section 7.

\textit{4) Sensitivity}:
Different from the above factors, which are calculated by RCOs, the sensitivity value is customized by each client.
%For instance, $C_i$ fetches the weighted opinions from RCO and applies the personalized sensitivity value.
Such a factor, denoted as $\theta_i$ and within [0, 1], indicates the acceptability level of the client for receiving an unsatisfactory service.
The rationale behind sensitivity is that some AIGC tasks are significant (e.g., the 3D avatar generations in Metaverse, which directly determine the immersiveness) or urgent (e.g., the real-time VR rendering).
The failure of these critical tasks is less tolerable so that the clients can set a high $\theta_i$.

\subsubsection{Opinion Fusion and Reputation}
Up till now, $C_i$ has its own opinion to $M_j$, i.e., $\omega_{ij}^{loc}$, and can fetch the latest reference opinions to $M_j$ from all the other clients, i.e., $\omega_{kj}^{loc}$ ($\forall k \in \mathbb{C}$ and $k \neq i$).
Then, it conducts the following steps to fuse the local and reference opinions and acquire the final reputation.
Firstly, the average of the calibrated reference opinions is calculated.
Denote the weighting factors for $\omega_{kj}^{loc}$ as $\Omega_{kj} := [\mathcal{F}_{kj}^a, \mathcal{F}_{kj}^r, \mathcal{W}_{kj}]$. 
Then, the averaged reference opinion $\omega_{*j}^{ref}$ := \{$s_{*j}^{ref}, u_{*j}^{ref}, c_{*j}^{ref}$\} can be defined as
\begin{equation}
        \left\{
\begin{aligned}
s_{*j}^{ref} & = (1-\theta_i) \frac{\sum_{k\in\mathbb{C}}||\mu_\Omega\odot\Omega_{kj}||_2\,s_{kj}^{loc}}{\sum_{k\in\mathbb{C}}||\mu_\Omega\odot\Omega_{kj}||_2} \\
u_{*j}^{ref} & = \theta_i \frac{\sum_{k\in\mathbb{C}}||\mu_\Omega\odot\Omega_{kj}||_2\,u_{kj}^{loc}}{\sum_{k\in\mathbb{C}}||\mu_\Omega\odot\Omega_{kj}||_2} \\
c_{*j}^{ref} & = \frac{\sum_{k\in\mathbb{C}}||\mu_\Omega\odot\Omega_{kj}||_2\,c_{kj}^{loc}}{\sum_{k\in\mathbb{C}}||\mu_\Omega\odot\Omega_{kj}||_2},
\end{aligned}
\right.
\end{equation}
where $||\cdot||_2$ is the 2-norm, $\odot$ represents the Hadamard product operation \cite{QoE1}, $\mu_\Omega$ := ($\mu_1$, $\mu_2, \mu_3$) are the weights of $\mathcal{F}_{kj}^a$, $\mathcal{F}_{kj}^r$, and $\mathcal{W}_{kj}$, respectively.
Then, $C_i$ fuses $\omega_{ij}^{loc}$ and $\omega_{*j}^{ref}$ and calculate the final opinion $\omega_{ij}$ := \{$s_{ij}^{fin}, u_{ij}^{fin}, c_{ij}^{fin}$\}, defined as
\begin{equation}
       \left\{
\begin{aligned}
s_{ij}^{fin} & = \frac{s_{ij}^{loc}c_{*j}^{ref} + s_{*j}^{ref}c_{ij}^{loc}}{c_{ij}^{loc}+c_{*j}^{ref} - c_{ij}^{loc}c_{*j}^{ref}} \\
u_{ij}^{fin} & = \frac{u_{ij}^{loc}c_{*j}^{ref} + u_{*j}^{ref}c_{ij}^{loc}}{c_{ij}^{loc}+c_{*j}^{ref} - c_{ij}^{loc}c_{*j}^{ref}}  \\
c_{ij}^{fin} & = \frac{c_{*j}^{ref}c_{ij}^{loc}}{c_{ij}^{loc}+c_{*j}^{ref} - c_{ij}^{loc}c_{*j}^{ref}}. 
\end{aligned}
\right. 
\end{equation}
Finally, the reputation of $M_j$, from $C_i$ perspective, can be calculated by
\begin{equation}
    \mathcal{R}_{ij} = s_{ij}^{fin} + \gamma c_{ij}^{fin}.
\end{equation}
The reputation value is adopted to reflect $\mathrm{OS}^2\mathrm{A}_S$.
From the above description, we can observe that such a scheme can accommodate heterogeneous clients with diverse standards, QoE models, and preferences.
Moreover, it is compatible with multiple AIGC modalities.
\begin{figure}[tbp]
\centerline{\includegraphics[width=0.9\columnwidth]{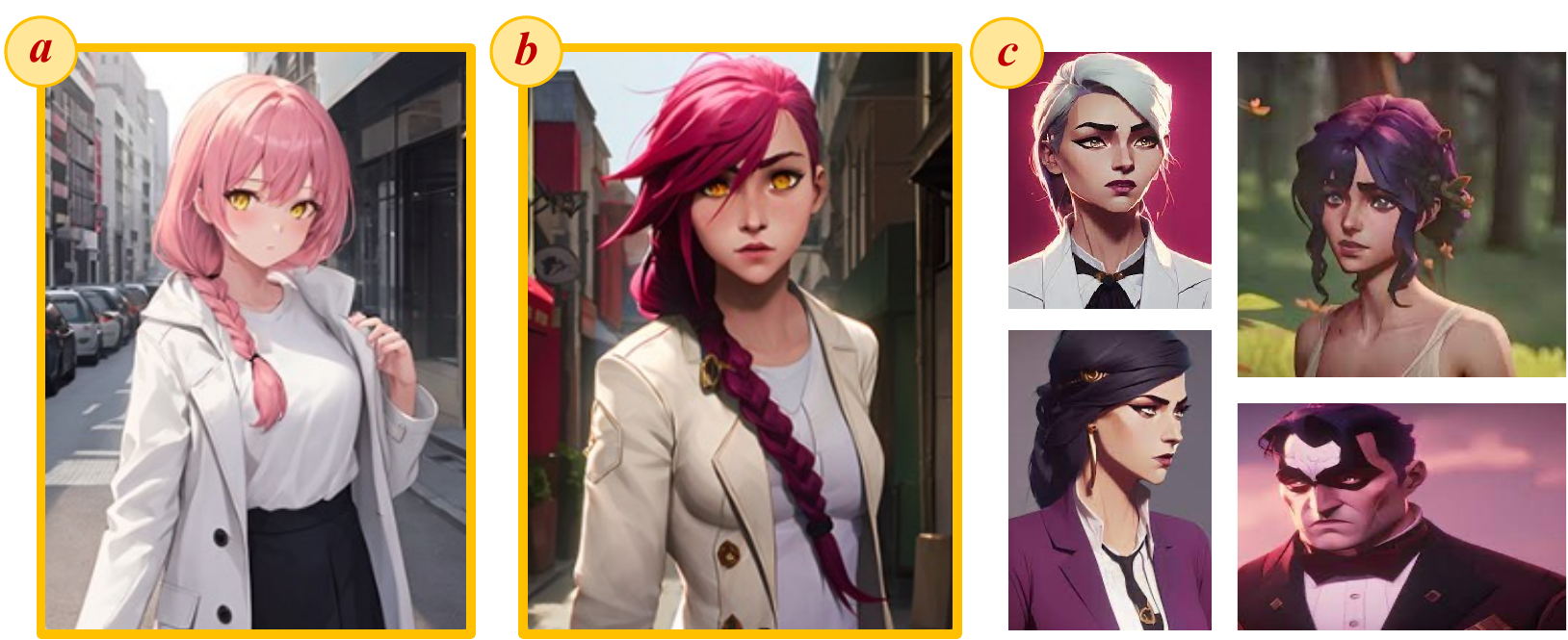}}
\caption{The perceivable experience of AIGC output quality. Images (\textit{a}) and (\textit{b}) are trained on \textit{AnyLoRA-Checkpoint} model using the same prompt: \textit{\textbf{Arcane style}, 1 girl, pink hair, long hair, one braid, white shirt, coat, yellow eyes, looking at viewer, city street}. However, the model for Image  (\textit{b}) is fine-tuned by \textit{Arcane Style} LoRA. Image (\textit{c}) shows some references of \textit{Arcane style} from Google. This example is from \cite{LoRa}.}
\label{qoe1}
\end{figure}

\subsubsection{Discussion: Rationale of Reputation} 
In this part, we discuss why $\mathrm{OS}^2\mathrm{A}_S$ is meaningful and can be reflected by reputation.
Recall that the judgment of AIGC outputs is implicit and might be affected by many latent factors.
However, clients can still perceive the difference between high- and low-quality outputs.
Accordingly, MASPs also have strategies to improve $\mathrm{OS}^2\mathrm{A}_S$, such as fine-tuning.
Fig. \ref{qoe1} illustrates an example, where two images about \textit{an Arcane style girl} are generated by the same basic AIGC model and with the same text prompt.
However, the model for Fig. \ref{qoe1}(b) is fine-tuned by the \textit{Arcane Style} Low-rank Adaptation (LoRA) \cite{ttt}.
Obviously, the resulting image exhibits more features of the Arcane style, such as a high nose bridge and strong facial contour.
This example demonstrates that honest clients will not randomly give their judgment (i.e., whether the AIGC outputs are satisfying or not).
Instead, their judgment is semantically meaningful.
On the other hand, if the MASP keeps providing customized AIGC services, e.g., downloading various LoRA and fine-tuning every generated AIGC output following clients' preferences, it can maintain a high reputation accordingly.

\section{$\mathrm{OS}^2\mathrm{A}$ on ProSecutor: Two-Phase Interaction for Mobile AIGC}
In this section, we deploy $\mathrm{OS}^2\mathrm{A}$ on \textsf{ProSecutor} and propose a two-phase scheme for clients to interact with MASPs.
Firstly, the most reliable MASP with the highest reputation can be selected.
Then, the clients can clarify the expected $\mathrm{OS}^2\mathrm{A}_O$ and utilize a contract theoretic approach to optimize the payment scheme.

\subsection{MASP Selection by Reputation}
Recall that the resources (e.g., $\Bar{c}$ and $\mathcal{F}_e$) that MASPs invest in the AIGC inference are unobserved; the clients can calculate the actual $\mathrm{OS}^2\mathrm{A}_O$ only after the entire service process.
In contrast, the reputation-based $\mathrm{OS}^2\mathrm{A}_S$ is calculated by the historical performance of each MASP, which is prior knowledge of all the clients.
Hence, we let clients adopt $\mathrm{OS}^2\mathrm{A}_S$ as the standard for MASP selection in the AIGC market with heterogeneous candidates.

The entire reputation scheme illustrated in Section 5 is deployed in \textsf{ProSecutor}, where RCOs hold the latest RT copy.
Hence, clients can easily acquire the reputation of all available MASPs via querying nearby RCOs.
The higher reputation value indicates that the MASP is more likely to generate high-quality outputs.
As shown in \textbf{Algorithm 2}, the entire MASP selection follows a multi-way handshake protocol.
We consider client $C_i$ and MASP $M_j$.
Initially, $C_i$ pings to $M_j$ and sends the service requests, including the service requirements, e.g., task description, expected $\mathrm{OS}^2\mathrm{A}_O$, and payment scheme.
If $M_j$ rejects the request, $C_i$ contacts the next candidate following the descending order of reputation.
Otherwise, $M_j$ returns an acknowledgment, which confirms the details by appending its digital signature.
Then, $C_i$ starts the second round of handshake, in which it sends the prompts for guiding AIGC inference to $M_j$.
Such prompts are encrypted by $M_j$'s public key, preventing it from being released to malicious third parties.
Likewise, $M_j$ returns the second-round acknowledgment and then formally performs the AIGC inference. 
%Note that for multimodal AIGC with complicated prompts, such as music composing that requires the demo audio and textual description, $C_i$ can save them on the storage chain and send the keys to $M_j$.
Next, we explain the design of the payment scheme in mobile AIGC using contract theory. 

\subsection{Contract Theoretic Payment Scheme}
Since MASPs invest tremendous physical resources in AIGC inference, they should be incentivized to ensure the network's economic equilibrium.
However, producing an appropriate payment scheme is challenging, especially in public mobile networks without mutual trust.
The most critical concern is moral hazard, i.e., MASPs may actually invest less resources than what they committed to obtain illegal revenue.
To this end, we employ contract theory \cite{moralhazard, Contract2} to optimize the payment scheme design.

\subsubsection{Problem Formulation}
We set the payment scheme as \textit{post-paid}, i.e., clients pay the service fee after receiving the AIGC outputs.
Compared with pre-paid/subscription, clients can be encouraged to participate in mobile AIGC by post-paid due to the lack of trustworthiness in mobile environments.
Note that such a setting will not lower MASPs' profits since the fee-ownership transfer protocol mentioned in Section 4.3 can defend the repudiation of both sides.
Furthermore, the amount of the AIGC service fee is designed as
\begin{equation}
    I_M = \mathcal{F}_s(\mathcal{D}_t) + \mu_s\cdot\mathrm{OS}^2\mathrm{A},
\end{equation}
where $\mathcal{F}_s(\mathcal{D}_t)$ is the fixed reward to MASP and is related to the difficulty of the AIGC tasks.
In contrast, $\mu_s$ means the bonus for unit $\mathrm{OS}^2\mathrm{A}$ value. 
In this way, the MASPs can be encouraged to invest as many resources as possible to pursue higher revenue.
Accordingly, if they misbehave, driven by moral hazard, the resulting service fee drops.

\textit{1) Utility of Clients:} 
Each client can gain profits from the AIGC outputs in multiple ways, e.g., offering Metaverse services for downstream markets using the brought AIGC avatars or directly reselling the outputs to others.
Nonetheless, the physical resources leased by MASPs should be rewarded.
Hence, the utility of the client is defined as the subtraction between these two parts, i.e.
\begin{equation}
    U_\mathrm{C} = \Big[(1 + \Bar{\epsilon})\,I_M\frac{\mathrm{OS}^2\mathrm{A}}{\mathbb{E}(\mathrm{OS}^2\mathrm{A})}\Big] - I_M,
\end{equation}
where the first term means the revenue it can acquire from the AIGC output with the specific $\mathrm{OS}^2\mathrm{A}$ value.
Hence, $\Bar{\epsilon}$ and $\mathbb{E}(\mathrm{OS}^2\mathrm{A})$ indicate the average \textit{return on investment} and the originally expected $\mathrm{OS}^2\mathrm{A}$ of clients, respectively.
We can observe that the higher the actual $\mathrm{OS}^2\mathrm{A}$ value, the higher the revenues that clients can earn.
Meanwhile, the value of $I_M$ increases accordingly.

\textit{2) Utility of MASP:}
MASPs gain profits by conducting AIGC inferences for clients.
Their costs come from two sources.
Firstly, the AIGC inferences consume enormous computation resources and power.
Additionally, MASPs should pay the transaction fee for channel establishment if the channels between them and the target clients have not been built.   
Therefore, the utility of MASPs can be expressed as
\begin{equation}
    U_\mathrm{SP} = I_M - \Big[\mathbb{E}(\mathcal{F}_e) + r\,\Bar{c}\,\mathcal{T}_2\Big],
\end{equation}
where $r$ is the unit cost of $\Bar{c}$.
The notation $\mathbb{E}(\mathcal{F}_e)$ indicates MASPs' expectation for the channel usage charge. 
For the sake of simplicity, we consider all MASPs to be conservative, i.e., they believe that the current round of interaction is the last AIGC service that they provide for this client. 
Therefore, assuming that it is the $\zeta^{th}$ round of interaction, $\mathbb{E}(\mathcal{F}_e)$ can be expressed as $\mathcal{F}_e/\zeta$.

\textit{3) Contract Modelling:}
The payment scheme is realized as contracts produced by clients to the selected MASPs.
Each contract has two items \{$\mu_s$, $\mathcal{F}_s$\}, which specify the basic fee and bonus rate set by the clients, respectively.
To design the optimal contract, we intend to maximize the clients' utility while providing the MASPs with the necessary incentives to agree on the contract.
Such an optimization problem can be expressed as 
\begin{equation}
\begin{array}{cc}
\mathop{\textnormal{max}}\limits_{\mu_s, \mathcal{F}_s, \Bar{c}, \mathcal{F}_e} & U_{\mathrm{C}}\left(\mu_s, \mathcal{F}_s, \Bar{c}, \mathcal{F}_e\right) \\
\text { s.t. } & \left\{\begin{array}{l}
\{\Bar{c}^{*}, \mathcal{F}_{e}^*\} \in \mathop{\textnormal{max}}\limits_{\Bar{c}, \mathcal{F}_e} U_{\mathrm{SP}}\left(\mu_s, \mathcal{F}_s, \Bar{c}, \mathcal{F}_e\right) \\
U_{\mathrm{SP}}\left(\mu_s, \mathcal{F}_s, \Bar{c}^*, \mathcal{F}_{e}^*\right) \geq U_{\mathrm{th}}^{\mathrm{SP}}.
\end{array}\right.
\end{array}
\end{equation}
The first constraint is Incentive Compatibility (IC), which emphasizes that the resources actually invested by MASPs should lead to the highest $U_\mathrm{SP}$.
Note that IC reflects the intrinsic pursuit of MASPs to maximize their revenue \cite{Contract1}.
The second constraint, i.e., Individual Rationality (IR), requires that the upper bound of $U_{\mathrm{SP}}$ should exceed the threshold since MASPs might be reluctant to accept service requests if the resulting revenue is too low \cite{Contract1}.

\subsubsection{Contract Finding by Diffusion-DRL}
%Traditionally, the optimal contract can be solved by convex optimization techniques \cite{MWSL}.
To acquire optimized contracts in complicated mobile AIGC scenarios, we adopt diffusion Deep Reinforcement Learning (DRL) \cite{du2023generative}.
%In this paper, we adopt the diffusion-DRL in \cite{du2023generative}.
Following the DRL paradigm, firstly, the \textit{action} space is composed of different contract designs, i.e., the set of \{$\mu_s$, $\mathcal{F}_s$\}.
The \textit{reward} is $U_\mathrm{C}$ corresponding to each contract design if both IC and IR are satisfied, and \textbf{-500} otherwise.
Finally, the \textit{state} is expressed as $e := \{\mathcal{S}_p, \Bar{b}, \mathcal{D}_t, \mathcal{S}_b, |\mathbb{C}|, \Bar{k}, p_b, E(\mathcal{L}), \lambda, \Bar{\epsilon}, r, \zeta\}$.
With all the components being configured, the diffusion-DRL utilities Double Q-Networks (DQNs) \cite{DQN} to explore the optimal contracts.
Particularly, the actor networks are implemented by diffusion paradigm.
In this case, leaning policies, i.e., mapping each state to the optimal action, can be abstracted as denoising from randomness to the optimal contracts guided by the Q-value.
% Add two more sentences with strong information
Note that a UNet realizes such denoising with learnable parameters and diffusion scheme \cite{du2023generative}.
Meanwhile, evaluation networks are employed to minimize the difference between the predicted Q-value by the current network and the real Q-value.
Compared with the conventional DRL approaches, introducing the diffusion process can further strengthen the model's exploration ability and adaptability to complex environments.
More architectural and algorithmic details are in \cite{du2023generative}.

%In \textsf{ProSecutor}, the diffusion-DRL model is trained and written into a specific smart contract.
%Once a client intends to produce the optimal contract, it can call the smart contract with $e$.
%Then, the execution results, i.e., the optimal contract, can be returned to it.
\begin{figure*}[tbp]
\centerline{\includegraphics[width=2\columnwidth]{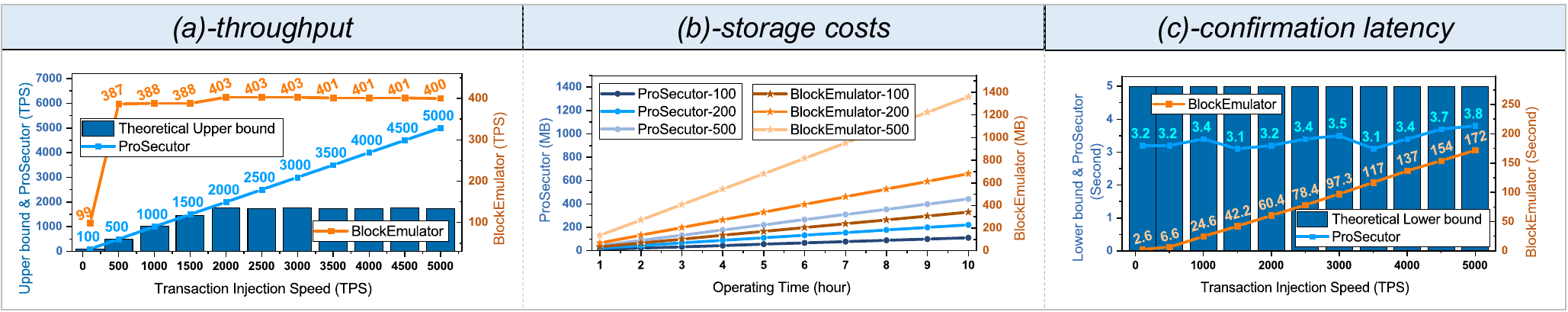}}
\caption{The performance evaluation of \textsf{ProSecutor}.}
\label{performance}
\end{figure*}

\section{Implementation and Evaluation}
In this section, we elaborate on the implementation and evaluation.
Firstly, we demonstrate the implementation of \textsf{ProSecutor} system.
Then, we conduct extensive experiments to validate the architectural superiority of \textsf{ProSecutor} and prove that \textsf{ProSecutor} can effectively achieve all the design goals.

\subsection{Implementation and Experimental Setup}
\;\;\;\;\;\textbf{Implementation}. We implement the prototype of \textsf{ProSecutor} atop \textit{BlockEmulator}\footnote{https://www.blockemulator.com/}, which is an open-source repository for building customized blockchain systems.
Specifically, we extend BlockEmulator by implementing the reputation roll-up and duplex transfer channels illustrated in Section 4. 
Afterward, we deploy the $\mathrm{OS}^2\mathrm{A}$ framework, reputation-based MASP selection, and contract theoretic payment scheme on \textsf{ProSecutor}.

\textbf{Testbed}. To construct the mobile AIGC market, we build an eight-node \textsf{ProSecutor} cluster.
The MASPs are served by Apple MacBook Pro with one 2.3 GHz 8-Core Intel Core i9 CPU and AMD Radeon Pro 5500M GPU. 
The mobile AIGC inferences are supported by \textit{Draw Things}, an iOS-oriented text-to-image AIGC application based on \textit{Stable Diffusion} model.
To fully simulate the real-world scenario, we feed \textsf{ProSecutor} with one million historical Ethereum transactions in 2022.
Finally, the baseline is the original version of BlockEmulator, which reproduces the traditional blockchain architecture, such as Bitcoin and Ethereum.

\textbf{Questions}. Through the experiments, we intend to answer the following research questions:
\begin{itemize}
    \item Can reputation roll-up and duplex and transfer channels reduce the blockchain overhead, making \textsf{ProSecutor} adapt to the mobile environment (for \textbf{G1})?
    \item Can \textsf{ProSecutor}, with the $\mathrm{OS}^2\mathrm{A}$ framework and mechanisms deployed on it, effectively realize all the functional goals, i.e., efficient MASP selection, payment scheme optimization, and atomic fee-ownership transfer (for \textbf{G2, G3, G4})?
    \item Can \textsf{ProSecutor} defend the attacks and threats in the mobile AIGC, e.g., forking ledgers, forging reputation, and breaking transfers?
\end{itemize}

\subsection{ProSecutor Performance Evaluation}
Firstly, we evaluate the performance and resource efficiency of \textsf{ProSecutor}.
We consider the following metrics.
\begin{itemize}
    \item \textit{Throughput}: The average number of transactions that can be processed in one second. The unit is Transaction per Second (TPS).
    \item \textit{Storage Costs}: The size of the entire blockchain ledger. The units are bytes or MegaBytes (MB).
    \item \textit{Confirmation Latency}: The time from when a transfer operation is submitted to when it is registered on the ledger. The unit is second.
\end{itemize}
Figs. \ref{performance}(a)-(c) show the throughput, storage costs, and confirmation latency of BlockEmulator and \textsf{ProSecutor}, respectively.
Specifically, we change the workload of the mobile AIGC market by adjusting the speed of transaction injection.
For blockchain configurations, the block generation interval is five seconds.
Each anchor chain and roll-up block can carry up to 2000 and 500 transactions, respectively. 

\textit{1) Throughput}:
Fig. \ref{performance}(a) illustrates that the throughput of BlockEmulator stalls at 387 TPS after the workload reaches 500 TPS. 
%Without roll-up, all the pending transactions are validated, confirmed by the consensus mechanism, and synchronized to the entire network.
To further explore the throughput upper bound of the traditional blockchain architecture, we enable BlockEmulator to generate one new block every second and use a latency-free internal network for message synchronization.
As illustrated by the blue bars in Fig. \ref{performance}(a), the throughput cannot exceed 1740 TPS even in such an ideal environment.
The limited throughput greatly hinders the system's scalability in adapting to large-scale AIGC markets with huge reputation lists.
With reputation roll-up, however, all \texttt{Opinion\_\!Update} transactions can be compressed and offloaded from the anchor chain.
Moreover, multiple RCOs can work in parallel, each of which manages the reputation of a subset of MASPs.
Hence, the throughput for reputation processing can increase linearly with the increasing network scale, supporting frequent reputation updates while saving the valuable capacity of the anchor chain.

\textit{2) Storage Costs}:
Fig. \ref{performance}(b) shows the storage costs with and without reputation roll-up when the workload is 100 TPS, 200 TPS, and 500 TPS.
In BlockEmulator, each block header and transaction occupy 120 and 99 bytes, respectively.
% Recall that the block generation takes 5 seconds and each block can carry at most 2000 transactions.
As a result, the size of the anchor chain ledger grows 34.07 MB/hour, 68.06 MB/hour, and 136.03 MB/hour under the workload of 100 TPS, 200 TPS, and 500 TPS.
Such storage costs are unaffordable for numerous mobile devices with limited storage capacity, such as smartphones and laptops.
Contributed to reputation roll-up, \textsf{ProSecutor} can offload all the \texttt{Opinion\_\!Update} transactions from the anchor chain, only maintaining the 32-byte transaction hashes.
Consequently, the storage costs can be reduced by at most 67.5\%. 
%These transactions are saved only in the RCOs, serving by high-end mobile edge servers.
Furthermore, RCOs can upload historical reputations to the storage chain and thus further reduce local storage costs as long as they can fetch the integral transaction information for reputation tracing.

\textit{3) Confirmation Latency}:
Fig. \ref{performance}(c) illustrates the confirmation latency of BlockEmulator and \textsf{ProSecutor}, with the increasing workload from 100 TPS to 5000 TPS.
Firstly, blue bars show the theoretical lower bound, which is five seconds, i.e., the block generation interval.
We can observe that BlockEmulator suffers from increasing latency since more transactions are kept in the transaction pool when the speed of transaction injection exceeds the maximum throughput.
Such a high latency prevents clients from efficiently utilizing the mobile AIGC services.
In contrast, assisted by duplex transfer channels, \textsf{ProSecutor} enables each client-MASP pair to perform ownership-fee transfers locally without queuing in the transaction pool.
Consequently, the confirmation latency is only determined by the channel capability for completing the procedure elaborated in Section 4.3, which is around 3.4 seconds in our testbed.
\begin{figure*}[htbp]
\centerline{\includegraphics[width=2\columnwidth]{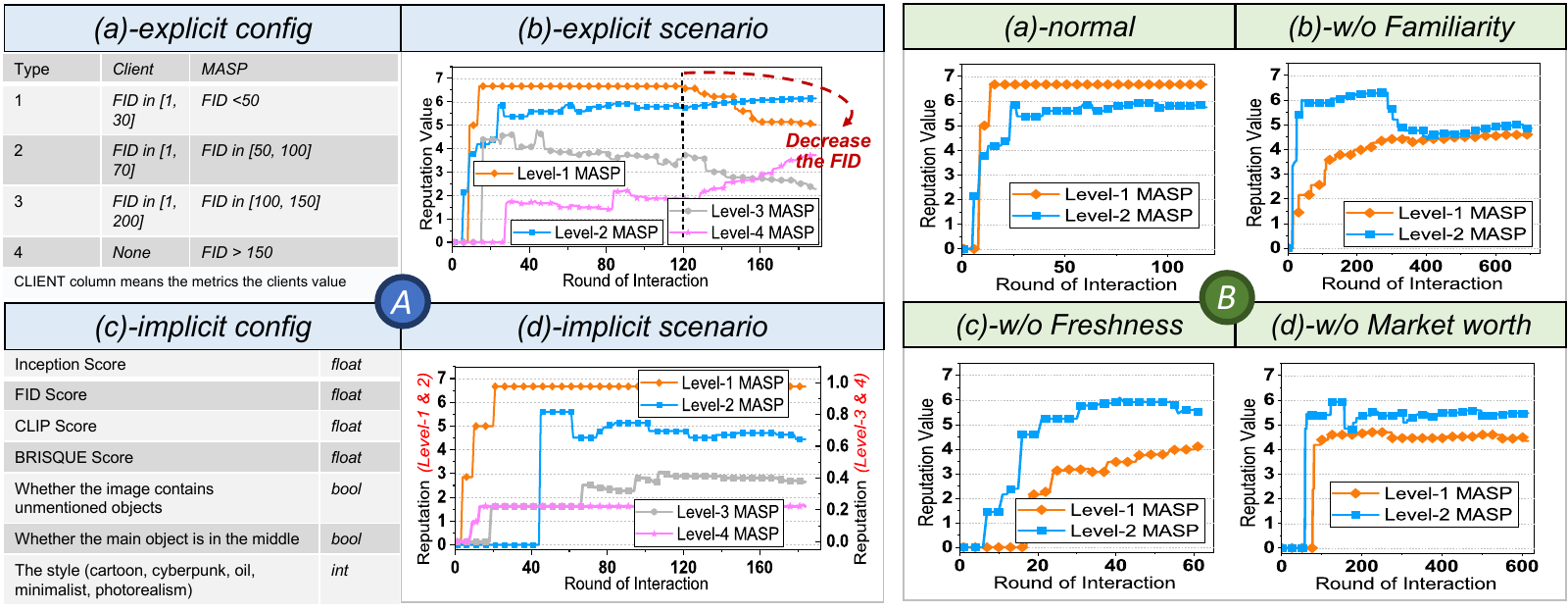}}
\caption{The inspection of MWSL-based reputation scheme for MASP selection. FID, SI, CLIP, and BRISQUE are general image quality assessment KPIs \cite{kastryulin2022pytorch}. The images with different KPI scores can be generated by first generating an original image and then adding noise/blur.}
\vspace{-0.3cm}
\label{qoe2}
\end{figure*}

\begin{figure}[tbp]
\centerline{\includegraphics[width=0.97\columnwidth]{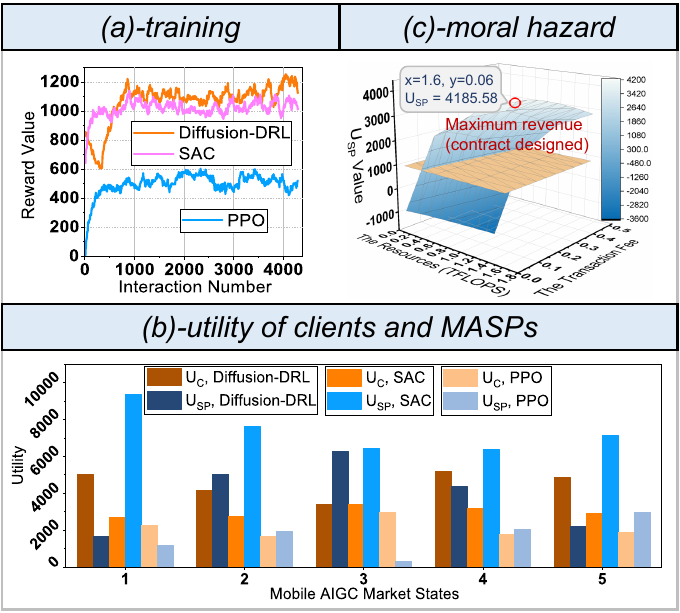}}
\caption{The training curve, utility, and defensive performance of contract theoretic payment scheme.}
\label{drl}
\end{figure}

\subsection{Investigation of Functional Goals}
The above experiments prove the architectural superiority of \textsf{ProSecutor} for adapting to mobile environments.
Next, we test whether \textsf{ProSecutor} can achieve the aforementioned three functional design goals, using the proposed on-chain mechanisms (including $\mathrm{OS}^2\mathrm{A}$ framework, contract theoretic payment scheme, and duplex transfer channels).

\subsubsection{Reputation-based MASP selection}
In this part, we evaluate the effectiveness of the proposed $\mathrm{OS}^2\mathrm{A}$ framework for MASP selection.
Recall that $\mathrm{OS}^2\mathrm{A}$ is measured by the MWSL-based reputation, allowing clients to express their subjective feelings via personalized models and standards.
We model these clients using two strategies: forming explicit and implicit scenarios.

\textit{1) Explicit Scenario}:
As shown in Fig. \ref{qoe2}-A(a), the explicit scenario contains four types of clients, which evaluate the AIGC images from Fréchet inception distance (FID) \cite{kastryulin2022pytorch}, but with different levels of strictness.
The larger the number of indexes, the higher the requirement for FID.
We further divide all the MASPs into four levels, with diverse abilities in terms of FID, also illustrated in Fig. \ref{qoe2}-A(a).
Then, we initialize such a mobile AIGC market, in which each client randomly selects the first MASP and then performs MASP selection following \textbf{Algorithm 2}.
The average reputation of each type of MASP is depicted in Fig. \ref{qoe2}-A(b).
From this figure, we can observe that reputation is semantically meaningful since the reputation values of MASPs strictly align with their capability levels.
Moreover, from the $120^{th}$ round, we intentionally decrease the FID values of level-1 and level-3 MASPs.
Accordingly, their reputation values drop gradually due to the increasing proportion of negative opinions.
Therefore, we can conclude that the reputation-based $\mathrm{OS}^2\mathrm{A}_S$ can effectively reflect the performance and behaviors of the MASPs. 

\textit{2) Implicit Scenario}:
To further simulate the practical mobile AIGC market, we consider the following implicit scenario.
Specifically, we use ChatGPT (empowered by GPT-4 model) to simulate the heterogeneous clients with diverse and latent QoE models.
As shown in Fig. \ref{qoe2}-A(c), given that ChatGPT cannot directly understand images, we evaluate each AIGC image from seven aspects.
Then, we let ChatGPT act as four types of clients, each of which generates a customized model to fuse a subset of metrics and sets a personalized strictness level to determine whether the AIGC service experience is positive or negative.
From Fig. \ref{qoe2}-A(d), we can observe that the reputation values are meaningful even in such a complicated scenario.
Moreover, the MASPs with higher capacity and providing customized services can gain higher reputations, even if the clients evaluate the AIGC services from different aspects and use diverse models.
In contrast, the traditional modality-specific models \cite{du2023generative} cannot accommodate these heterogeneous clients.

\textit{3) Ablation Study}:
Moreover, we conduct an ablation study to prove the effectiveness of the factors used for calibrating reference opinions.
Firstly, Fig. \ref{qoe2}-B(a) shows the reputation trend with all the factors.
Then, we disable the \textit{Familiarity} factor, in which case the attackers can launch the flooding attack \cite{8651742}.
To simulate it, we employ numerous clients to submit massive fake opinions, thereby increasing the reputation of the Level-2 MASPs.
As shown in Fig. \ref{qoe2}-B(b), Level-2 MASPs outperform Level-1 MASPs at the beginning.
However, the flooding opinions usually have a small market worth; otherwise, they are easy to detect and mark, which may lead to token freezing.
Therefore, after honest clients reach consistency on the Level-2 MASPs' actual reputation, the effect of flooding opinions gradually drops because of the \textit{Market worth} factor.
Next, we disable the \textit{Freshness} factor.
In this case, the system is vulnerable to long-range attacks \cite{8651742}, in which the attackers first forge several high-value opinions and thus gain great weight forever.
As shown in Fig. \ref{qoe2}-B(c), the reputation of Level-2 MASP exhibits a U-shape, with each uptick corresponding to a high-value opinion.
Since no more such opinions after the $30^{th}$ round, the increment stalls, while the effect caused by them can last for a long time.
Finally, we disable the \textit{Market worth} factor, in which case the attackers can launch dusting attacks by submitting massive opinions.
As shown in Fig. \ref{qoe2}-B(d), the reputation of Level-2 MASPs increases dramatically since massive high-value opinions are sent in the initial stage.
Hence, the attackers can obtain the highest familiarity and freshness levels, thus tempering the reputation of the target MASPs.

\subsubsection{Contract Theoretic Payment Scheme}
To optimize the contract design, we train the diffusion-DRL model \cite{du2023generative}, whose training curve is shown in Fig. \ref{drl}(a).
Note that we use two conventional DRL methods, namely Proximal Policy Optimization (PPO) and Soft Actor-Critic (SAC), as the baselines.
From Fig. \ref{drl}(a), we can observe that the diffusion-DRL converges faster and achieves a higher reward.
The reasons are two-fold: 1) the multiple-step diffusion keeps refining the model parameters to optimize the contract design, and 2) the long-term dependence in the complex mobile AIGC environments can be learned since the contract design is generated by steps rather than one-time in PPO and SAC.
Using the well-trained model for inferences, Fig. \ref{drl}(b) further shows the utility of clients by following the optimal contracts in different AIGC market states. 
We can observe that our method maintains the highest utility in all the tested situations, ensuring a stable revenue for the clients.
Additionally, the IC and IR constraints are satisfied, ensuring that the MASPs are willing to sign the contracts.
In contrast, PPO fails to reach the IR constraints in one state.
Finally, Fig. \ref{drl}(c) evaluates the defensive performance of our contract theoretic payment scheme against moral hazard.
We can find that regardless of how the MASP changes the contract parameters, their utility decreases.
Therefore, it will strictly follow the contract to maximize its utility, and the threats from the moral hazard can be circumvented.
\begin{figure}[tbp]
\centerline{\includegraphics[width=0.98\columnwidth]{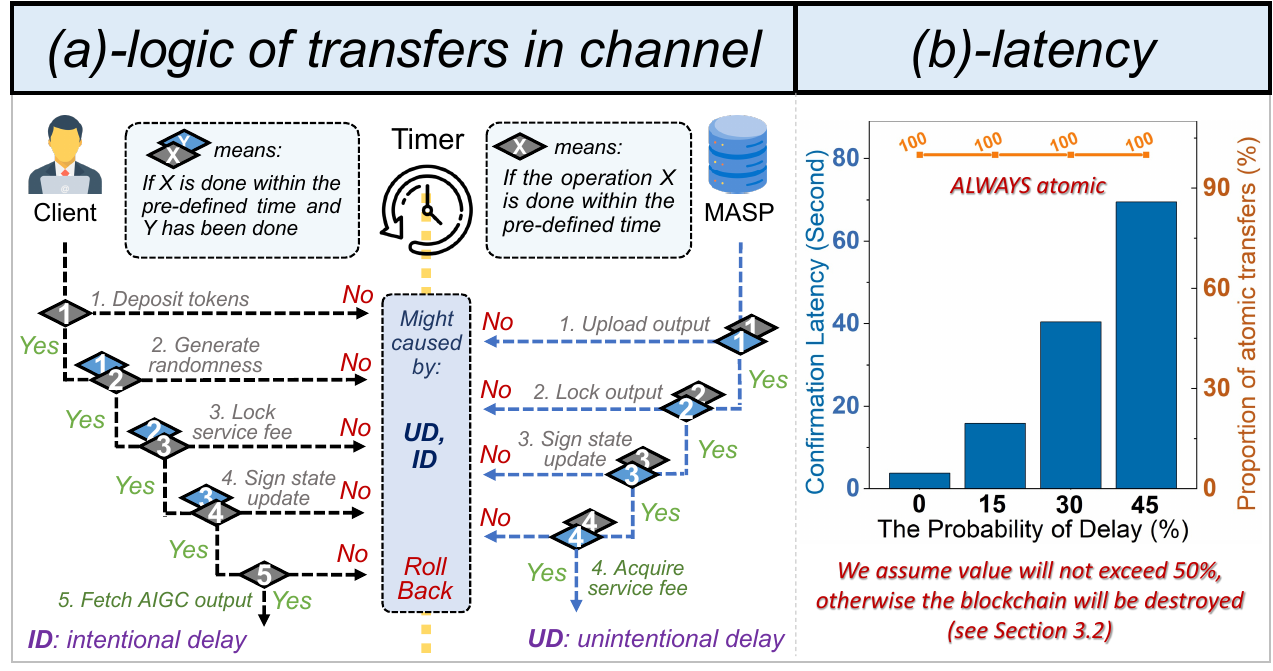}}
\caption{The atomicity of fee-ownership transfer.}
\label{transfer}
\end{figure}

\subsubsection{Atomic Fee-Ownership Transfers}
The atomicity of the fee-ownership transfer is protected by the smart contract, whose logic is presented in Fig. \ref{transfer}(a).
We can observe that only if the current instruction is executed correctly, as well as the other side performs as promised, can the program proceed to the next step.
If all the instructions are finished normally, the transfer succeeds.
Otherwise, the roll-back will be triggered without any immediate states.
The transfer contract is deployed on the anchor chain and instantiated by the transfer channels. 
All the operations (e.g., instruction fetches and executions) are automatically conducted when the pre-defined conditions are met, which guarantees that malicious attackers cannot manipulate the transfer process.

Although unable to tamper with the smart contracts, dishonest participants might disturb the normal contract execution by intentionally delaying the instruction executions. 
For instance, in Fig. \ref{transfer}(a), if the client continuously submits fake signatures of its updated balance in Step 4, not only its Step 5 but also Step 4 of the MASP will be affected.
The atomicity and latency of transfers are shown in Fig. \ref{transfer}(b), where \textit{Probability of Delay} represents the probability that one instruction execution is delayed.
We can observe that the transfer atomicity level maintains at 100\% due to the protocol logic and the features of the smart contract.
However, the latency continues to increase, adversely affecting the system performance and the user experience.
To defend the intentional delay, in practice, we can assign each instruction execution a timer whose value is customized during the channel establishment.
If one instruction fails to be completed within the pre-defined time, the rollback operation will be triggered.
Furthermore, the clients and MASPs can configure the timer according to specific tasks and scenarios (e.g., the time sensitivity of the service and the trustworthiness between the two parties).

\begin{table*}[!ht]
\renewcommand{\multirowsetup}{\centering}
\caption{The security issues and the corresponding defenses.}
\footnotesize
\centering
\renewcommand\arraystretch{1.27}
\begin{tabular}{m{3.78cm}<{\centering}|c|m{9cm}}
\rowcolor[rgb]{0.90,0.90,0.90}
\toprule[1pt]
\hline
\rowcolor[rgb]{0.90,0.90,0.90}
\multicolumn{1}{m{3.8cm}<{\centering}|}{\cellcolor[rgb]{0.90,0.90,0.90} \textbf{Objectives}}&\textbf{Issue (Definition) }&\multicolumn{1}{m{9cm}<{\centering}}{\cellcolor[rgb]{0.90,0.90,0.90} \textbf{Defenses in \textsf{ProSecutor} (Description and Experiment)}}\\
\hline
\multirow{3}*{\textbf{Destroy the system}}
&\multicolumn{1}{m{4cm}|}{51\% attack (Section 3.2)} & The DPoS consensus mechanism\\ 
\cline{2-3}
&\multicolumn{1}{m{4cm}|}{Identity attack (Section 3.2)} &  The SHA-256 public-private key pair for identity, signature, encryption/decryption \\
\cline{2-3}
&\multicolumn{1}{m{4cm}|}{Spamming/DoS (Section 5.2.1)} &  The transaction fee scheme \\
\hline
\multirow{3}*{\textbf{Tamper the reputation}}
&\multicolumn{1}{m{4cm}|}{Flooding attack (Section 7.3.1)} & \multicolumn{1}{m{9cm}}{\textit{Familiarity} factor in the reputation scheme (Sections 5.3.2 and 7.3.1)}\\
\cline{2-3}
&\multicolumn{1}{m{4cm}|}{Long-range attack (Section 7.3.1)} & \multicolumn{1}{m{9cm}}{\textit{Freshness} factor in the reputation scheme (Sections 5.3.2 and 7.3.1)}\\
\cline{2-3}
&\multicolumn{1}{m{4cm}|}{Dusting attack (Section 7.3.1)} & \multicolumn{1}{m{9cm}}{\textit{Market worth} factor in the reputation scheme (Sections 5.3.2 and 7.3.1)}\\
\hline
\multirow{1}*{\textbf{Gain additional profit}}
&\multicolumn{1}{m{4cm}|}{Moral hazard (Section 3.2)} & \multicolumn{1}{m{9cm}}{Contract theoretic payment scheme (Sections 6.2 and 7.3.2).}\\
\hline
\multirow{1}*{\textbf{Interrupt the transfer}}
&\multicolumn{1}{m{4cm}|}{Intentional delay (Section 7.3.3)} & \multicolumn{1}{m{9cm}}{Timer for each instruction (both in Section 7.3.3).}\\
\hline
\bottomrule[1pt]
\end{tabular}
\end{table*}

\subsection{Security Analysis}
TABLE 2 summarizes the defenses of \textsf{ProSecutor} against potential security issues.
Firstly, the attacks aiming to destroy \textsf{ProSecutor} in terms of the consistency, identity, and liveness \cite{KAD} can be effectively defended by DPoS, SHA256-based cryptography scheme, and transaction fee scheme, respectively.
Since these defenses are theoretically proven \cite{8651742, Xupaper}, we omit the detailed explanations. 
Secondly, during the opinion collection and reputation calculation, the attackers can adopt diverse ways to tamper with the values.
Leveraging MWSL \cite{MWSL}, we calibrate the reputation by \textit{Familiarity}, \textit{Freshness}, and \textit{Market worth} factors, which are proven to defend flooding, long-range, and dusting attacks effectively.
Furthermore, the contract theoretic payment scheme and duplex transfer channels are proven to defend the moral hazard and repudiation, respectively.

% Note that the IEEE does not put floats in the very first column
% - or typically anywhere on the first page for that matter. Also,
% in-text middle ("here") positioning is typically not used, but it
% is allowed and encouraged for Computer Society conferences (but
% not Computer Society journals). Most IEEE journals/conferences use
% top floats exclusively. 
% Note that, LaTeX2e, unlike IEEE journals/conferences, places
% footnotes above bottom floats. This can be corrected via the
% \fnbelowfloat command of the stfloats package.

\section{Conclusion}
In this paper, we presented \textsf{ProSecutor}, the first blockchain system for protecting mobile AIGC.
Specifically, contributed to roll-up and layer-2 channels, \textsf{ProSecutor} achieves high performance and resource efficiency to adapt to the mobile environment.
With \textsf{ProSecutor}, we first proposed the atomic fee-ownership transfer protocol, defending the repudiation in trustless mobile networks.
Then, we presented a novel framework for QoE modeling in mobile AIGC, called $\mathrm{OS}^2\mathrm{A}$.
By fusing objective service KPIs and reputation-based subjective experience, $\mathrm{OS}^2\mathrm{A}$ can efficiently evaluate the AIGC services and guide clients to select the MASP with the highest probability of providing satisfying AIGC outputs.
Moreover, we utilized contract theory to help clients optimize payment scheme design and circumvent the moral hazard.
Extensive experiments demonstrated the effectiveness and validity of \textsf{ProSecutor}.

\textbf{Future Work}: For the future work, \textsf{ProSecutor} can be improved from the following aspects. First, the underlying architecture of \textsf{ProSecutor} can be updated to sharding blockchain, further improving the throughput and resource efficiency. Secondly, more objective KPIs oriented to Mobile AIGC can be integrated into $\mathrm{OS}^2\mathrm{A}$, thus increasing the effectiveness of AIGC service assessment. Finally, the human feedback can be incorporated into the diffusion-DRL, thus making the decisions align with user preferences.

% or


\ifCLASSOPTIONcaptionsoff
  \newpage
\fi

\bibliographystyle{IEEEtran}
\bibliography{parma}
\vfill



%\bibliographystyle{IEEEtran}


%\vfill


\end{document}